\documentclass{aa}  

\usepackage{graphicx}
\usepackage[varg]{txfonts}
\usepackage{lscape}
\usepackage{gensymb}
\usepackage{braket}
\usepackage{footmisc}

\newcommand{\mearth}{M_\oplus}
\newcommand{\rearth}{R_{\rm \oplus}}
\newcommand{\msun}{M_\odot}
\newcommand{\rsun}{R_{\rm \odot}}
\newcommand{\lsun}{L_\odot}
\newcommand{\rhosun}{\rho_\odot}

\newcommand{\rjup}{R_\mathrm{jup}}

\newcommand{\logg}{\log g}

\def\ms{\hbox{\,m\,s$^{-1}$}}         
\def\gcc{\hbox{\,g\,cm$^{-3}$}}         
\def\m2s2{\hbox{\,m$^{2}$\,s$^{-2}$}} 
\def\kms{\hbox{\,km\,s$^{-1}$}}       
\def\sini{\hbox{sin\,$i$}}      

\begin{document}

   \title{So close, so different: characterization of the K2-36 planetary system with HARPS-N}


   \author{M. Damasso \inst{\ref{oato}} 
   \and L.~Zeng\inst{\ref{harvplan}}
   \and L.~Malavolta\inst{\ref{oapd},\ref{pduni}}
   \and A. Mayo\inst{\ref{berk}}   
   \and A. Sozzetti\inst{\ref{oato}} 
   \and A.~Mortier \inst{\ref{camb}}
   \and L. A.~Buchhave\inst{\ref{dtuspace}}
   \and A.~Vanderburg\inst{\ref{unitx},\ref{cfa},\ref{sagan}}  
   \and M. Lopez-Morales\inst{\ref{cfa}}
   \and A.~S.~Bonomo\inst{\ref{oato}}
   \and A.~C.~Cameron\inst{\ref{supa}}
   \and A.~Coffinet\inst{\ref{obsge}}
   \and P.~Figueira\inst{\ref{esochile},\ref{uporto}}
   \and D. W.~Latham\inst{\ref{cfa}}
   \and M.~Mayor \inst{\ref{obsge}}
   \and E. ~Molinari \inst{\ref{oaca}}
   \and F.~Pepe \inst{\ref{obsge}}
   \and D. F.~Phillips\inst{\ref{cfa}}
   \and E.~Poretti\inst{\ref{tng},\ref{brera}}
   \and K.~Rice\inst{\ref{ediuni}}
   \and S.~Udry \inst{\ref{obsge}}
   \and C.A.~Watson \inst{\ref{belfast}}
}
   \institute{INAF - Osservatorio Astrofisico di Torino, Via Osservatorio 20, I-10025 Pino Torinese, Italy\label{oato}\\
   \email{mario.damasso@inaf.it}
   \and Department of Earth and Planetary Sciences, Harvard University, Cambridge, MA 02138 \label{harvplan}
   \and INAF - Osservatorio Astronomico di Padova, Vicolo dell'Osservatorio 5, 35122 Padova, Italy\label{oapd}
   \and Dipartimento di Fisica e Astronomia ``Galileo Galilei", Universita'di Padova, Vicolo dell'Osservatorio 3, 35122 Padova, Italy\label{pduni}
   \and UC Berkeley Astronomy Department, Berkeley, CA 94720-3411 \label{berk}
   \and Astrophysics Group, Cavendish Laboratory, University of Cambridge, JJ Thomson Avenue, Cambridge CB3 0HE, UK \label{camb}
   \and DTU Space, National Space Institute, Technical University of Denmark, Elektrovej 327, DK-2800 Lyngby, Denmark \label{dtuspace}
   \and Department of Astronomy, The University of Texas at Austin, 2515 Speedway, Stop C1400, Austin, TX 78712\label{unitx}
   \and Harvard-Smithsonian Center for Astrophysics, 60 Garden Street, Cambridge, MA 02138, USA\label{cfa}
   \and NASA Sagan Fellow\label{sagan}
   \and Centre for Exoplanet Science, SUPA, School of Physics and Astronomy, University of St Andrews, St Andrews KY16 9SS, UK\label{supa}
   \and Observatoire de Gen\`eve, Universit\'e de Gen\`eve, 51 ch. des Maillettes, 1290 Sauverny, Switzerland\label{obsge}  
   \and European Southern Observatory, Alonso de Cordova 3107, Vitacura, Santiago, Chile \label{esochile}
   \and Instituto de Astrof\'isica e Ci\^encias do Espa\c{c}o, Universidade do Porto, CAUP, Rua das Estrelas, 4150-762 Porto, Portugal\label{uporto}
  \and INAF – Osservatorio Astronomico di Cagliari, Via della Scienza 5 - 09047 Selargius CA, Italy \label{oaca}
   \and INAF - Fundaci\'on Galileo Galilei, Rambla Jos\'e Ana Fernandez P\'erez 7, 38712 Bre\~na Baja, Spain \label{tng}
   \and INAF-Osservatorio Astronomico di Brera, Via E. Bianchi 46, 23807 Merate, Italy \label{brera}
   \and Centre for Exoplanet Science, University of Edinburgh, Edinburgh, UK \label{ediuni} 
   \and Astrophysics Research Centre, School of Mathematics and Physics, Queen’s University Belfast, BT7 1NN, Belfast, UK \label{belfast} 
       }

   \date{}

 
  \abstract
   {K2-36 is a K dwarf orbited by two small ($R_{\rm b}=1.43\pm0.08$ $\rearth$ and $R_{\rm c}=3.2\pm0.3$ $\rearth$), close-in ($a_{\rm b}$=0.022 AU and $a_{\rm c}$=0.054 AU) transiting planets discovered by the Kepler/K2 space observatory. They are representatives of two distinct families of small planets ($R_{\rm p}$<4 $\rearth$) recently emerged from the analysis of Kepler data, with likely a different structure, composition and evolutionary pathways.}
   {We revise the fundamental stellar parameters and the sizes of the planets, and provide the first measurement of their masses and bulk densities, which we use to infer their structure and composition.}
   {We observed K2-36 with the HARPS-N spectrograph over $\sim$3.5 years, collecting 81 useful radial velocity measurements. The star is active, with evidence for increasing levels of magnetic activity during the observing time span. The radial velocity scatter is $\sim$17 \ms due to the stellar activity contribution, which is much larger that the semi-amplitudes of the planetary signals. We tested different methods for mitigating the stellar activity contribution to the radial velocity time variations and measuring the planet masses with good precision.}
   {We found that K2-36 is likely a $\sim$1 Gyr old system, and by treating the stellar activity through a Gaussian process regression, we measured the planet masses $m_{\rm b}$=3.9$\pm$1.1 $\mearth$ and $m_{\rm c}$=7.8$\pm$2.3 $\mearth$. The derived planet bulk densities $\rho_{\rm b}$=7.2$^{+2.5}_{-2.1}$ $\gcc$ and $\rho_{\rm c}$=1.3$^{+0.7}_{-0.5}$ $\gcc$ point out that K2-36\,b has a rocky, Earth-like composition, and K2-36\,c is a low-density sub-Neptune.}
  {Composed of two planets with similar orbital separations but different densities, K2-36 represents an optimal laboratory for testing the role of the atmospheric escape in driving the evolution of close-in, low-mass planets after $\sim$1 Gyr from their formation. Due to their similarities, we performed a preliminary comparative analysis between the systems K2-36 and Kepler-36, which we deem worthy of a more detailed investigation.}


   \maketitle
%

\section{Introduction}
Launched in 2009, the Kepler space observatory has discovered thousands of transiting extrasolar planets, and unveiled the huge diversity that characterizes the exoplanetary systems. Such large numbers of new worlds has enabled statistical studies to shed light on the different observed architectures, forming an important repository of data on the formation and evolutionary history of the planetary systems.

One striking feature that has emerged from observations is that the distribution of the radii of small ($R_{\rm p}$<4 $\rearth$), close-in planets is bi-modal (\citealt{owenwu13}, and the following studies by \citealt{fulton17,zeng17a,zeng17b}). By refining the stellar and the planet radii using parallaxes from \textit{Gaia}, \cite{fulton18} have better characterized these two quite distinct populations, identified as super-Earths and sub-Neptunes. They have determined that the centres of the two groups lie at 1.2-1.3 $\rearth$ and $\sim$ 2.4 $\rearth$, respectively, and were able to locate the radius gap that separates the two families as lying between $\sim$ 1.8-2 $\rearth$ (as also confirmed by \citealt{berger18}). The bi-modality in the size distribution is likely indicative of existing differences in the bulk compositions of the two classes of planets, which in turn could also be related to the formation history and the mass loss mechanisms in action during the system evolution, such as photo evaporation (e.g. \citealt{lopezfortney2013,owenwu17,vaneylen18}). 

There is ambiguity in the compositions of the sub-Neptune family, since they may be either gas dwarfs, with a rocky core surrounded by an H$_{\rm 2}$/He gaseous envelope, or water-worlds mainly composed of H$_{\rm 2}$O-dominated ices or/and fluids, or a combination of an ice mantle and an H$_{\rm 2}$/He gaseous envelope. According to Zeng et al. (2018, submitted) many of the 2-4 $\rearth$ planets could actually be water-worlds. Nonetheless, the key information that is still missing for the bulk of the small planets are their masses, which would allow for a determination of their average densities, necessary to constrain their compositions and internal structures. Therefore, the characterization of these small exoplanets through the measurement of their masses is fundamental to understanding the origins and diversity of planets in the Milky Way Galaxy.  

Within this framework, we present a characterization study of the K2-36 planetary system \citep{sinukoff16} based on radial velocities (RVs) extracted from spectra collected with the HARPS-N spectrograph. This system is of special interest in that the star K2-36 hosts two small, close-in transiting planets discovered by K2 that are exemplar members of the two planet families identified in the Kepler sample ($R_{\rm b}$ = 1.43$\pm$0.08 $R_{\rm \oplus}$ and $R_{\rm c}$ = 3.2$\pm$0.3 $R_{\rm \oplus}$). K2-36 offers the exciting opportunity to measure the mass of planets below and above the radius gap that belong to the same system, testing the hypothesis that the gap represents the transition between rocky planets and lower density bodies with enough volatiles to measurably change their bulk composition.
K2-36 is a magnetically active star with an RV scatter of $\sim$17 \ms, which is much larger than the expected semi-amplitude of the planetary signals on the basis of the observed planet radii. Thus, it represents a very challenging case study concerning the characterization of small-size, low-mass planets, even with transit ephemeris known to high precision.

The paper is organized as follows. In Section \ref{sect:dataset} we describe the photometric and spectroscopic datasets used in this work. In Section \ref{sect:stellarparam} we provide a new determination of the fundamental stellar parameters for K2-36, making use of \textit{Gaia} data, and in Section \ref{sect:lcanalisi} we present refined planetary parameters from a reanalysis of the K2 light curve. We discuss in Section \ref{sect:actindex} the results of the stellar activity analysis, and present in Section \ref{sect:rvanalysis} the measurements from RV data of the mass and bulk density of the two K2-36 planets. In Section \ref{sect:discussion} we discuss the implications of our results concerning the composition and bulk structure of the K2-36 planets. The main conclusions are outlined in Section \ref{sect:conclusion}.


\section{Dataset}
\label{sect:dataset}

\subsection{K2 photometry}
\begin{figure}
   \centering
   \includegraphics[width=\hsize]{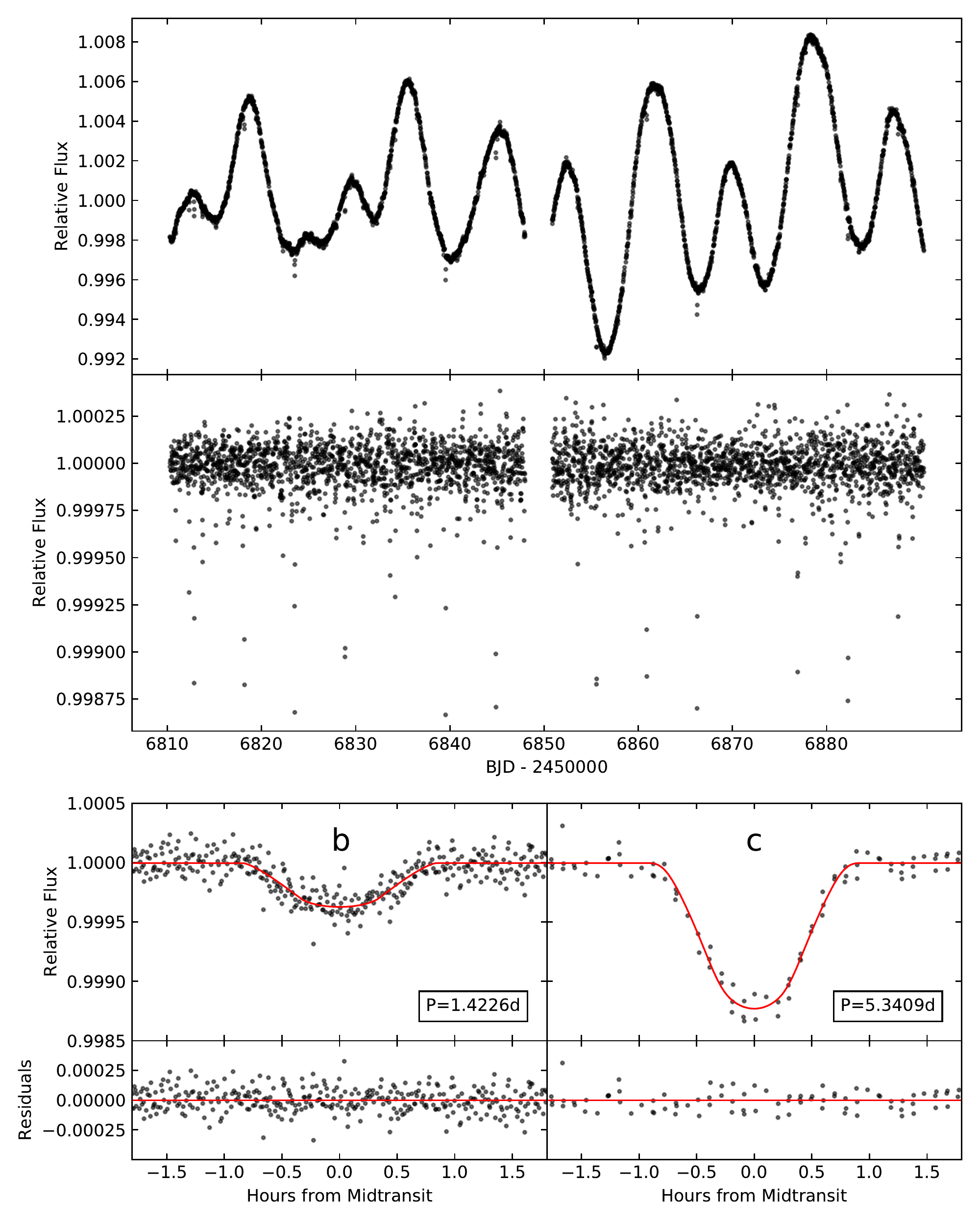}
      \caption{\textit{Upper panel}: K2-36 light curve observed by K2, showing the modulation induced by the stellar rotation. \textit{Middle plot}: K2-36 flattened light curve, with the rotation modulation filtered-out. The observed dimming events correspond to the transits of the two planets. \textit{Bottom panel}: transit light curves for K2-36\,b and K2-36\,c folded at their best-fit orbital periods (Table \ref{Table:planetfittransit}). The curves in red are our best-fit transit models, and the residuals are also shown.}
         \label{Fig:curvaluce}
   \end{figure}

K2-36 was observed by the K2 mission during Campaign 1 in 2014 (May 30-Aug 21).
In our work we used the 30 minute cadence K2SFF\footnote{https://archive.stsci.edu/prepds/k2sff/} light curve processed as described by \cite{vanderjohns14} and \cite{vander16}, and refined by simultaneously fitting for spacecraft systematics, stellar variability, and the planetary transits. The light curve is shown in the upper panel of Fig. \ref{Fig:curvaluce}. A gap is visible between epochs BJD 2\,456\,848 and 2\,456\,850.9, during which the observations were interrupted in order to downlink the data. During that time Kepler changed orientation, which also changed the heating due to the Sun and introduced an offset in the photometry. The middle panel in Fig. \ref{Fig:curvaluce} shows the flattened light curve corrected for the modulation due to the stellar rotation, and it is that we used to model the planetary transit signals. 

\subsection{HARPS-N spectroscopy}
We observed K2-36 over three seasons, between the end of January 2015 and the end of May 2018, with the high-resolution and stabilised spectrograph HARPS-N \citep{cosentino14} at the Telescopio Nazionale Galileo (TNG) on the island of La Palma. A total of 83 echelle spectra have been secured on fiber A, all with an exposure time of 1800 seconds. Their average signal-to-noise ratio S/N per pixel is 34, as measured at wavelength $\lambda\sim$550 nm. Two of them have S/N<15 (taken at epochs BJD 2\,457\,810.586681 and 2\,457\,890.464286), and they have been discarded from further analysis due to their low quality. Fiber B was used to collect sky spectra. The spectra were reduced with version 3.7 of the HARPS-N Data Reduction Software (DRS). 

We performed an analysis to identify the RV measurements potentially contaminated by moonlight. This check is particularly necessary for a target such as K2-36, which lies close to the ecliptic plane. The procedure adopted for this analysis is the same as the one described in Section 2.1 of \cite{malavolta17a}. We have found one potentially contaminated spectrum at epoch BJD 2\,457\,054.632798, but since our analysis is not conclusive we have decided to include it in the final dataset, which thus consists of 81 spectra.

\section{Stellar parameters}
\label{sect:stellarparam}

Adaptive optics imaging did not detect stellar companions near K2-36 \citep{sinukoff16}, and there is no evidence in the cross-correlation function (CCF) of the HARPS-N spectra for more than one component.
We derived new values for the fundamental stellar parameters from the analysis of HARPS-N spectra and using data from the \textit{Gaia} DR2 catalogue. The atmospheric parameters were first determined separately using three different algorithms. 

\textit{Equivalent widths}. In the classical curve-of-growth approach, we derive temperature T$_{\rm eff}$ and microturbulent velocity ${\xi }_{{\rm{t}}}$ by minimizing the trend of iron abundances (obtained from the equivalent width of each line) with respect to excitation potential and reduced equivalent width, respectively. The surface gravity $\logg$ is obtained by imposing the same average abundance from neutral and ionized iron lines. We used \texttt{ARESv2}\footnote{Available at http://www.astro.up.pt/$\sim$sousasag/ares/} \citep{sousa15} to measure the equivalent widths, and used \texttt{MOOG}\footnote{Available at http://www.as.utexas.edu/$\sim$chris/moog.html} \citep{sneden73} jointly with the \texttt{ATLAS9} grid of stellar model atmospheres from \cite{castelli04} to perform line analysis and spectrum synthesis, under the assumption of local thermodynamic equilibrium (LTE). We followed the prescription of \cite{andreasen17} and applied the gravity correction from \cite{mortier14}. The analysis was performed on the resulting co-addition of individual spectra. We get $T_{\rm eff}=4800\pm59$ K, $\logg=4.73\pm$0.15 (cgs), and [Fe/H]=-0.15$\pm$0.03 dex. 

\textit{Atmospheric Stellar Parameters from Cross-Correlation Functions (CCFpams)}. This technique is described in \cite{malavolta17b}, and the code is publicly available\footnote{https://github.com/LucaMalavolta/CCFpams}. We get $T_{\rm eff}=4841\pm37$ K, $\logg=4.60\pm$0.10 (cgs), and [Fe/H]=-0.19$\pm$0.04 dex. We report here the internal errors, which are likely underestimated. 

\textit{Stellar Parameter Classification (SPC)}. The SPC technique \citep{buchhave12,buchave14} was applied to 55 spectra with high S/N, and the weighted average of the individual spectroscopic analyses yielded stellar parameters of $T_{\rm eff}=4862\pm50$ K, $\logg=4.57\pm$0.10 (cgs), and [m/H]=-0.05$\pm$0.08 dex. For the projected rotational velocity we can reliably impose only the upper limit $v\sin i_{\rm \star}$<2 \kms.

\subsection{The age of K2-36}
Inferring the age of this planetary system is relevant for investigating its origin and evolution, but it is a difficult issue for a K dwarf when using only stellar evolution isochrones. Based on the rotational modulation observed in the $K2$ light curve (Sect. \ref{lcstellarrotation}) and on a rotation-age relationship (Scott Engle, private communication), the age of K2-36 is found to be 1.5$\pm$0.4 Gyr, which agrees with our measured average level of chromospheric activity $\braket{\log R\ensuremath{'}_{\rm HK}}$ = -4.50 dex.

Using the relation of \cite{mamajek08} based on gyrochronology (eqs. [12]-[14]), and adopting P$_{\rm rot}$=16.9$\pm$0.2 as the rotation period of K2-36 (Sect.  5.1), we derive an age of 1.09$\pm$0.13 Gyr, although the color index of K2-36 ($B-V$=0.96, corrected for the extinction $E(B-V)$=0.03 using the 3D Pan-STARRS 1 dust map of \citealt{green18}) is slightly out the range of validity for that calibration (0.5<$B-V$<0.9). This result is in agreement with the previous estimate.

We also carried out Galactic population assignment using the classification scheme by \cite{bensby03,bensby05}, the K2-36's systemic radial velocity from HARPS-N spectra, and \textit{Gaia} DR2 proper motion and parallax. We obtain a thick disk–to–thin disk probability ratio $thick/thin=0.1$, implying that K2-36 is significantly identified as a thin-disk object, presumably not very old. 

\begin{figure}
     \centering
     \includegraphics[width=\hsize]{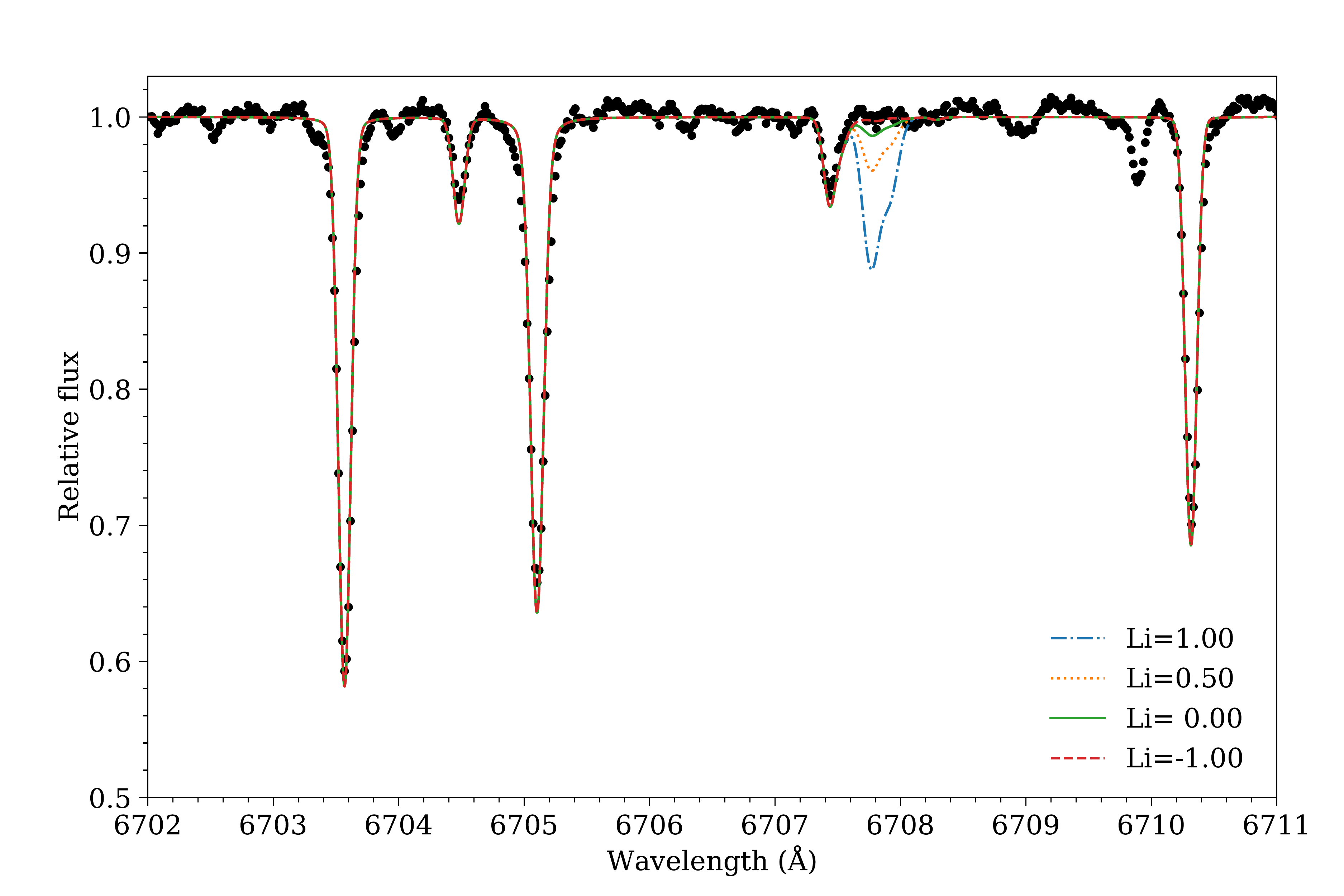}
     \includegraphics[width=\hsize]{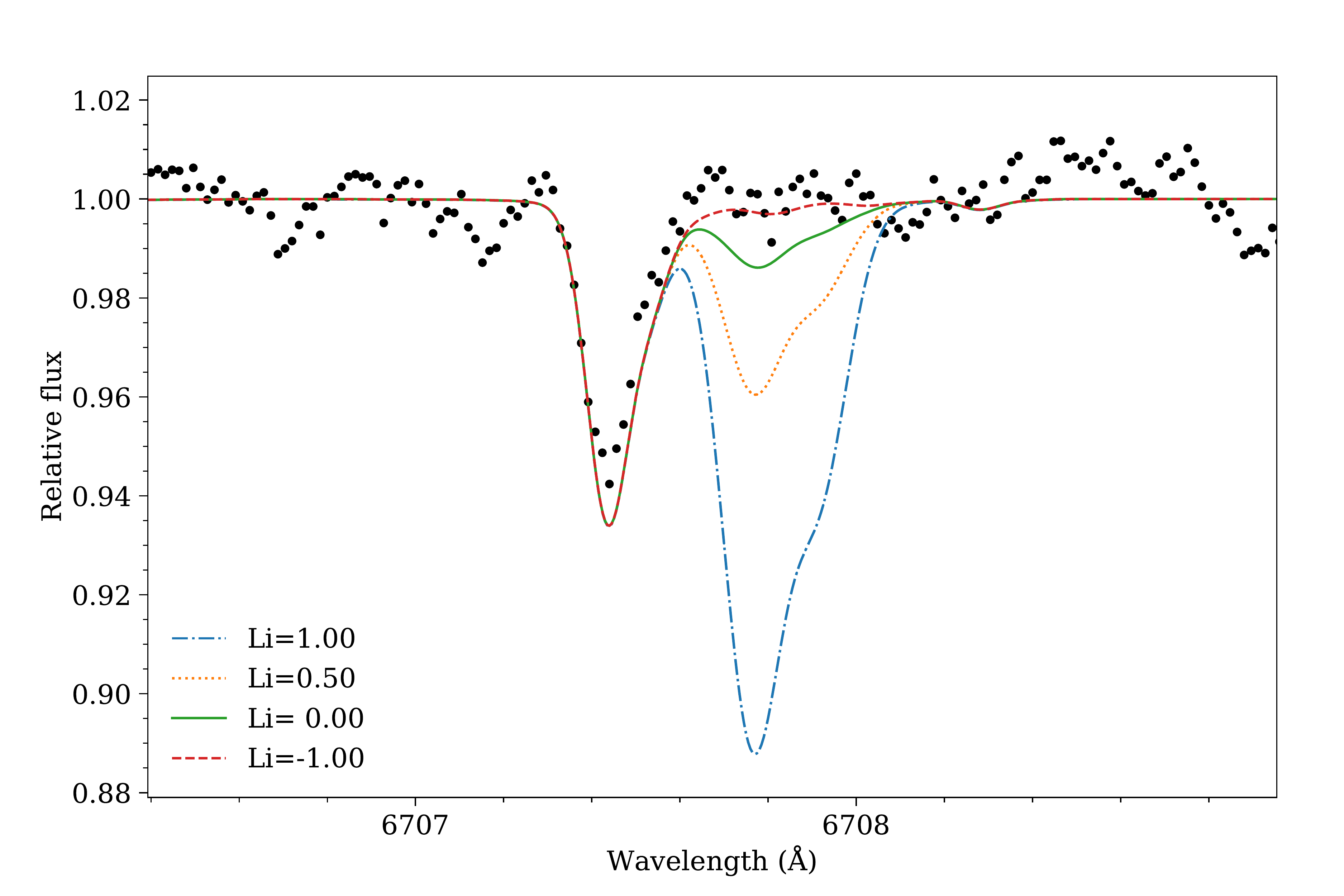}
      \caption{\textit{Top panel}: portion of the HARPS-N co-added spectrum of K2-36 containing
       the LiI line at 6707.8 \AA , compared to four different syntheses (lines of various colors and styles), each differing only for the assumed lithium abundance. \textit{Bottom panel}: zoom-in view of the region around 6707.8 \AA to enable easier comparison between the observed data and model spectra for low Li abundances.”}
         \label{Fig:litium}
   \end{figure}

Finally, we searched for evidence of the lithium absorption line at 6707.8 \AA\: (Fig. \ref{Fig:litium}). We compared the co-added spectrum with four models, each differing only in the assumed Li abundance. The lithium line is not detectable within the noise of the continuum, and we set an upper limit of $A$(Li)<0.0. The lithium depletion in K2-36 is comparable to, or lower than, that observed for stars of the same temperature and spectral type in 600-800 Myr-old open clusters with similar metallicity (e.g., \citealt{soderblom95,sestito05,somers14,brandt15}). We thus set an approximate lower limit for the age of K2-36 at $t=0.6$ Gyr.

\subsection{Age-constrained final set of stellar parameters}
We used our estimated age of K2-36 to derive a final set of fundamental stellar parameters, including the stellar mass and radius. To this purpose we have constrained the stellar age to be in the range of 600 Myr to 2 Gyr. The parameters were obtained with the package \texttt{isochrone} \citep{morton15}, following the prescriptions of \cite{malavolta18}.
The photospheric parameters of the star obtained with the three different techniques outlined earlier in this Section were used as priors together with the photometric magnitudes V, K, WISE1 and WISE2, and the parallax from \textit{Gaia}, within a Bayesian framework. The stellar evolution models Dartmouth \citep{Dotter2008} and MIST \citep{Dotter2016,Choi2016,Paxton2011} were used, thus a total of six different sets of posteriors for the fitted parameters have been derived. The final results are listed in Table \ref{Tab:starparam2} and represent the 50$^{\rm th}$, 15.86$^{\rm th}$ and 84.14$^{\rm th}$ percentiles (the last two used for defining the error bars) after combining all the posteriors together (32310 samples) to take into account the differences existing between the different stellar evolution models. Our model provides an estimate of the extinction $A_{\rm V}$ at the distance of the star, $A_{\rm V}$=0.09$^{+0.09}_{-0.07}$ mag. This value is consistent with zero and is in agreement with the results of the 3D dust mapping with Pan-STARRS 1.

\begin{table}
  \caption{Summary of the fundamental stellar parameters of K2-36.}
         \label{Tab:starparam2}
         \small
         \centering
   \begin{tabular}{l l l}
            \hline
            \noalign{\smallskip}
            Parameter  &  Value & Ref. \\
            \noalign{\smallskip}
            \hline
            \noalign{\smallskip}
            RA [ICRS J2015] & 11:17:47.78 & (1) \\
            \noalign{\smallskip}
            DEC [ICRS J2015] & +03:51:59.00 & (1) \\
            \noalign{\smallskip}
            $\mu_{\rm \alpha^{*}}$ [mas/year] &-20.521$\pm$0.076 & (1) \\
            \noalign{\smallskip}
            $\mu_{\rm \delta}$ [mas/year] & 26.194$\pm$0.058 & (1) \\
            \noalign{\smallskip}
            Parallax [mas] & 9.0832$\pm$0.0504 & (1) \\
            \noalign{\smallskip}
            $V$ [mag] & 11.803$\pm$0.030 & (2) \\
            \noalign{\smallskip}
            $B-V$ & 0.992$\pm$0.064 & (2) \\
            \noalign{\smallskip}
            $K_{\rm s}$ [mag] & 9.454$\pm$0.025 & (3) \\
            \noalign{\smallskip}
            WISE/W1 [mag] & 9.379$\pm$0.023 & (4) \\
            \noalign{\smallskip}
            WISE/W12 [mag] & 9.442$\pm$0.020 & (4) \\
            \noalign{\smallskip}
			Radius [$\rsun$] & 0.718$^{+0.008}_{-0.006}$ & (5a) \\ 
            & 0.74$\pm$0.04 & (6) \\
            & 0.75$\pm$0.02 & (7) \\
			\noalign{\smallskip}
            Mass [$\msun$] & 0.79$\pm$0.01 & (5a) \\
            & 0.80$\pm$0.04 & (6) \\
            & 0.80$^{+0.02}_{-0.03}$ & (7) \\
            \noalign{\smallskip}
			Density [$\rhosun$] & 2.12$\pm$0.04 & (5a) \\
            \noalign{\smallskip}
            $T_{\rm eff}$ [K] & 4916$^{+35}_{-37}$ & (5) \\ 
            & 4924$\pm$60 & (6) \\
            & 4944$\pm$50 & (7) \\
            \noalign{\smallskip}
            $\logg$ [cgs] & 4.621$^{+0.005}_{-0.004}$ & (5b) \\
            &  4.65$\pm$0.10 & (6) \\
            &  4.70$\pm$0.10 & (7) \\
            \noalign{\smallskip}
            $[\rm Fe/H]$ [dex] & -0.09$^{+0.06}_{-0.04}$ & (5) \\
            & -0.03$\pm$0.04 & (6) \\
            & -0.03$\pm$0.08 & (7) \\
            \noalign{\smallskip}
            Age [Gyr] & 1.4$^{+0.4}_{-0.5}$ & (5a) \\ 
            & 1.5$\pm$0.4 & (5c) \\
            & 1.1$\pm$0.2 & (5d) \\
            \noalign{\smallskip}
            $\log(L/\lsun$) & -0.57$\pm$0.01 & (5a) \\
            \noalign{\smallskip}
            $v\sini_{\rm \star}$ [\kms] & < 2 & (5) \\
            & 2$\pm$1 & (6) \\
            \noalign{\smallskip}
            $\braket{\log R\ensuremath{'}_{\rm HK}}$ [dex] & -4.50 & (5) \\
             \noalign{\smallskip}
            P$_{\rm rot}$ [days] & 16.9$\pm$0.2 & (5) \\
            \noalign{\smallskip}
            \hline
     \end{tabular}  
     \tablefoot{(1) \textit{Gaia} DR2 \citep{gaiacoll16,brown18}; (2) APASS DR9 \citep{apass15}; (3) 2MASS Catalog \citep{2mass06}; (4) WISE Catalog; (5) this work; (5a) this work, derived from isochrones by imposing an age in the range 600 Myr-2 Gyr; (5b) derived from isochrones, and not from spectral analysis; (5c) based on a rotation-age relationship, and consistent with the measured $\braket{\log R\ensuremath{'}_{\rm HK}}$; (5d) based on the activity-age calibration of \cite{mamajek08}; (6) \cite{sinukoff16}; (7) \cite{mayo18}.}
\end{table}

\section{Photometric transit analysis}
\label{sect:lcanalisi}

The transit analysis was done in the same manner as described in \cite{mayo18}. The two planets in the system were fit simultaneously using the \texttt{BATMAN} transit fitting package. The model consists of four global parameters (baseline flux level, a noise parameter, and a quadratic limb darkening law) as well as five parameters per planet (time of transit, period, $R_{\rm p}/R_{\rm \star}$, semi-major axis, and inclination). Our model also assumes zero eccentricity and non-interacting planets. The parameters were estimated using \texttt{emcee} \citep{foreman13}, a python package to perform Markov chain Monte Carlo (MCMC) analysis. Convergence was determined by requiring the Gelman-Rubin statistic \citep{gelman92} to be less than 1.1 for all parameters. The only difference between our analysis and that of \cite{mayo18} is that we included a prior on the stellar density $\rhosun$ in the light curve model fit. Specifically, at each step in the MCMC simulation we took the stellar density estimated from the transit fit at that step and applied a Gaussian prior to penalize values discrepant from our spectroscopic estimate of $\rhosun$. Table \ref{Table:planetfittransit} summarizes the fit results. We see that, while our results are consistent with those in \cite{mayo18}, the prior on the stellar density reduced the uncertainties in planetary radii, inclinations and transit durations. Planet radii are consistent within 1$\sigma$ with those derived by \cite{sinukoff16}, and have a similar precision.

Based on their radii, K2-36\,b and K2-36\,c are each representative of one of the two planet populations in the radius distribution, below $\sim$1.8 $\rearth$ and above $\sim$2 $\rearth$, as characterized by \cite{fulton18}. 
We note that the mutual inclination between the orbital planes of the planets ($\Delta I\leq 3 \degree$) is in agreement with the results of \cite{dai18} based on the analysis of Kepler/K2 multi-planet systems (not including K2-36). They found that when the innermost planet is closer to the host star than $a/R_{\rm \star}$=5, $\Delta I$ can be likely greater than $\sim5\degree$, and this effect could also depend on the orbital period ratio, with pairs with $P_{\rm c}/P_{\rm b}\geq$5-6 showing larger $\Delta I$. K2-36\,b has $a/R_{\rm \star}$=6.6, the period ratio is $P_{\rm c}/P_{\rm b}\sim$3.8, and the measured $\Delta I$ is very close to the dispersion of 2.0$\pm$0.1 degrees they found for multi-planet systems where the innermost planet has 5 < $a/R_{\rm \star}$ < 12. This suggests that K2-36 likely belongs to a group of systems that evolved through a mechanism which did not excite the inclination of the innermost planet, therefore without resulting in different orbital architectures. 

\section{Stellar activity analysis}
\subsection{Stellar rotation from K2 light curve}
\label{lcstellarrotation}
\begin{figure}
   \centering
   \includegraphics[width=\hsize]{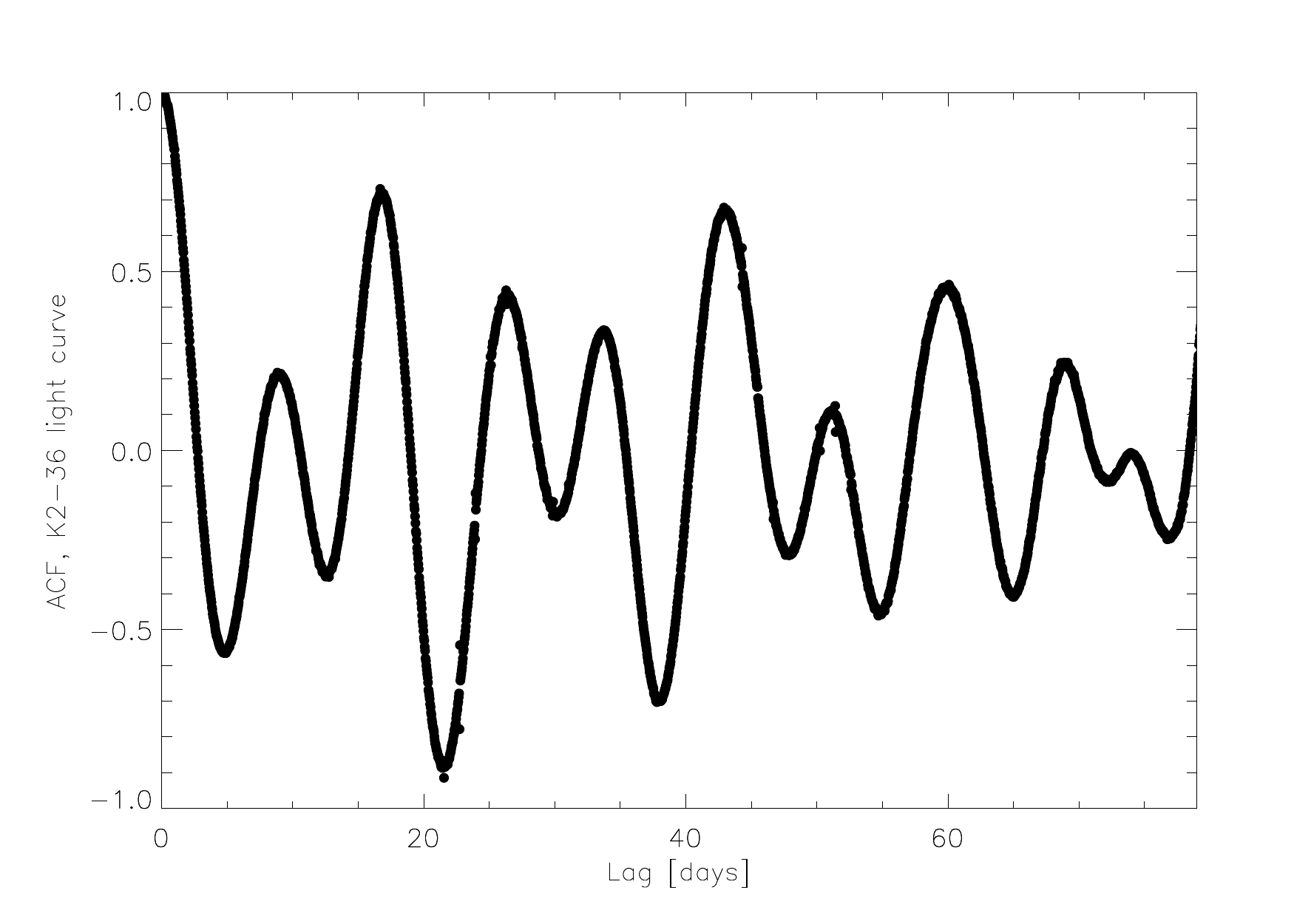}
      \caption{Autocorrelation function of the K2-36 light curve shown in Fig. \ref{Fig:curvaluce}.}
         \label{Fig:acf}
\end{figure}

Looking at the light curve of K2-36 (Fig. \ref{Fig:curvaluce}, upper panel), the modulation induced by the stellar rotation can be clearly seen. The data before the gap show a two-maxima pattern that repeats every $\sim$17 days, suggesting that this could be the actual stellar rotation period $P_{\rm rot}$. 
The scatter of the data before the gap is $\sigma_{\rm phot}$=2.4$\%$, which increases to $\sigma_{\rm phot}$=3.9$\%$ for data collected after the gap. A positive trend is seen in the second batch of data, that suggests that the active regions on the stellar photosphere are evolving with a timescale comparable with the stellar rotation period. However, we note that the light curve extracted with the alternative pipeline \texttt{EVEREST} \citep{luger16} shows a weaker trend (light curve not shown here), implying that it could be not entirely astrophysical in origin. The highest increase in the relative flux of the light curve maxima is limited to $\sim$0.2$\%$, and it is observed between the last two rotation cycles.    
To determine $P_{\rm rot}$ we computed the autocorrelation function (ACF) of the light curve using the DCF method by \cite{edelson88} (Fig. \ref{Fig:acf}). The ACF has been explored up to the 80-day time span. The highest ACF peak occurs at $\sim$16.7 days. We modelled this peak with a Gaussian profile\footnote{using \texttt{LMFIT}, a Non-Linear Least-Squares Minimization and Curve-Fitting for Python.} in order to determine the error on $P_{\rm rot}$, that we assume to be equal to the Gaussian RMS width. We get $P_{\rm rot}$=16.7$\pm$1.9 days. 

Since the light curve shows evidence for quasi-periodic variations on a timescale of the order of $P_{\rm rot}$, this makes it interesting to analyse the data using a Gaussian process (GP) regression as done, e.g., by \cite{haywood14}, \cite{cloutier17}, or \cite{angus14} to model space-based photometry, adopting a quasi-periodic kernel defined by the covariance matrix

\begin{eqnarray} \label{eq:1}
K(t, t^{\prime}) = h^2\cdot\exp\bigg[-\frac{(t-t^{\prime})^2}{2\lambda^2} - \frac{sin^{2}\left(\dfrac{\pi(t-t^{\prime})}{\theta}\right)}{2w^2}\bigg] + \nonumber \\
+\, \sigma^{2}_{\rm phot, K2}(t)\cdot\delta_{t,t^{\prime}}
\end{eqnarray} \\
where $t$ and $t^{\prime}$ represent two different epochs. The four hyperparameters are $h$, which represents the amplitude of the correlations; $\theta$, which parametrizes the rotation period of the star; $w$, which describes the level of high-frequency variation within a complete stellar rotation; and $\lambda$, which represents the decay timescale of the correlations and can be physically related to the active region lifetimes. The flux error at time \textit{t} is indicated by $\sigma_{\rm phot, K2}(t)$, and $\delta_{t,t^{\prime}}$ is the Kronecker delta. We introduced a free parameter to model the offset due to the gap in the K2 observations. Since the computing time for a GP regression scales as N$^{\rm 3}$, where N is the number of data points, we analysed the 6-hr binned light curve instead of the full dataset.

\begin{figure}
   \centering
   \includegraphics[width=\hsize]{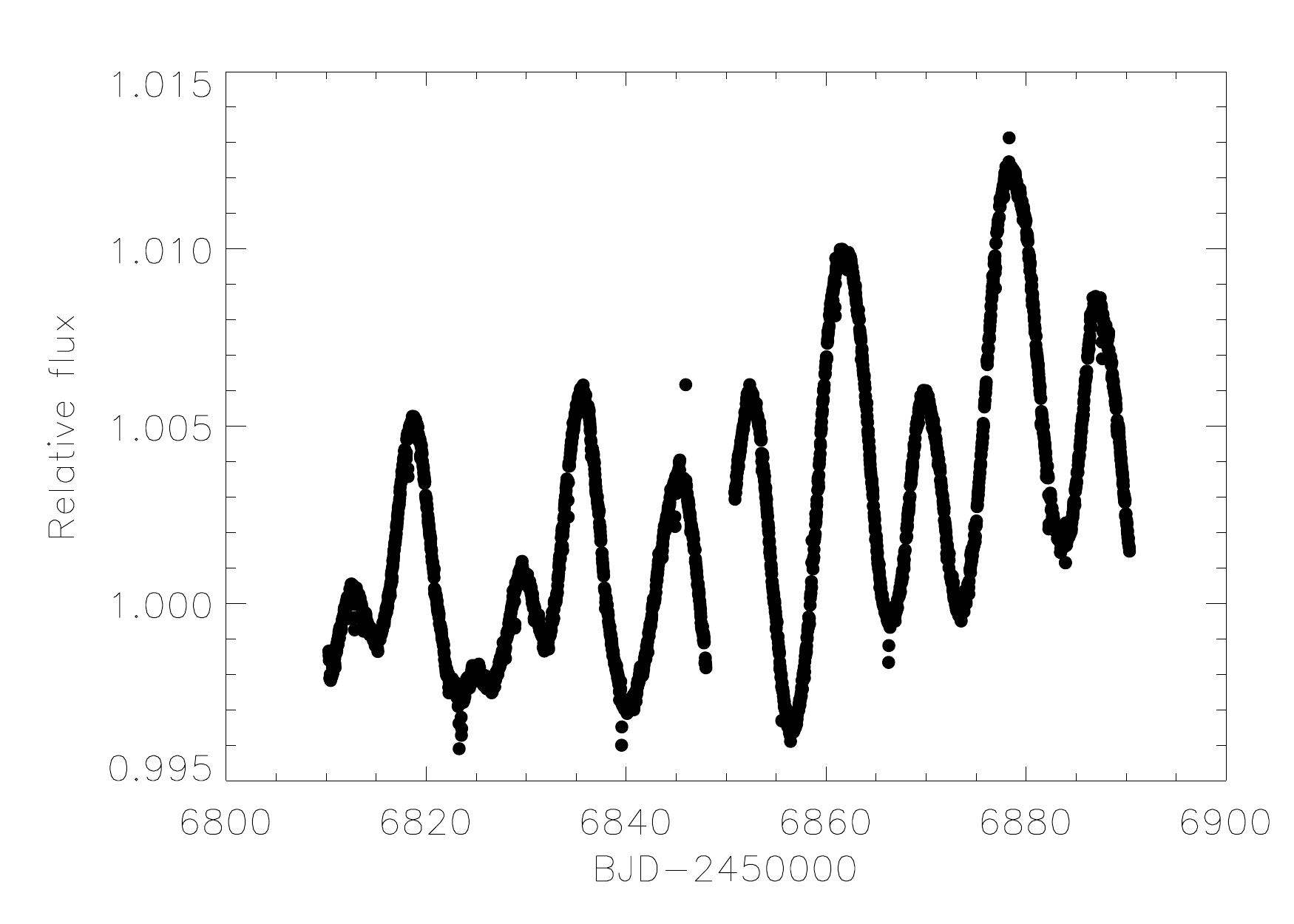}
      \caption{K2-36 light curve corrected for the offset introduced by the gap in the K2 observations (see Fig. \ref{Fig:curvaluce} for comparison). An offset of $\sim0.36\%$ has been determined through a Gaussian process regression, as described in Section \ref{sect:lcanalisi}. }
         \label{Fig:lccorrected}
\end{figure}

All the GP analyses presented in this work (see also Sect. \ref{sect:actindex} and \ref{sect:rvanalysis}) were carried out using the publicly available Monte Carlo sampler and Bayesian inference tool \texttt{MultiNestv3.10} (e.g. \citealt{feroz13}), through the \texttt{pyMultiNest} wrapper \citep{buchner14}, by adopting 800 live points and a sampling efficiency of 0.5. All the logarithms of the Bayesian evidence ($\ln\mathcal{Z}$) mentioned in our work were calculated by MultiNest. The GP component of our code is the publicly available \texttt{GEORGEv0.2.1} python module \citep{ambika14}. 
The results of the analysis are summarized in Table \ref{Tab:actindgp}. The rotation period, that we left free to vary between 0 and 20 days, is consistent with, but more precise than, the ACF estimate ($P_{\rm rot}$=16.9$\pm$0.2 days). The best-fit value for the evolution timescale of the active regions is well constrained ($\lambda=106_{\rm -11}^{\rm +14}$ days) and of the order of a few rotation periods. The best-fit value for the offset is 0.0036$\pm$0.0004, and Fig. \ref{Fig:lccorrected} shows the light curve corrected for the offset. Within the framework of a quasi-periodic model, the corrected light curve shows that the active regions on one hemisphere are evolving faster than those on the opposite side of the stellar disk, because the relative height of one maximum has increased by $\sim1.2\%$ over the observation time span.  

Using our measure for the stellar radius and the upper limit for $v\sini_{\rm \star}$ (which coincides with the best-fit value of \citealt{sinukoff16}), the expected maximum rotation period is $P_{\rm rot}=17.7$ days. 
This result implies that the projected inclination of the spin axis is $i_{\rm \star}$$\sim$90$\degree$.

Since the star is active, we searched for flares in the K2 light curve, without finding any clear evidence of large events.

\subsection{Stellar activity spectroscopic diagnostics}
\label{sect:actindex}
\begin{figure*}
   \centering
   \includegraphics[width=17cm]{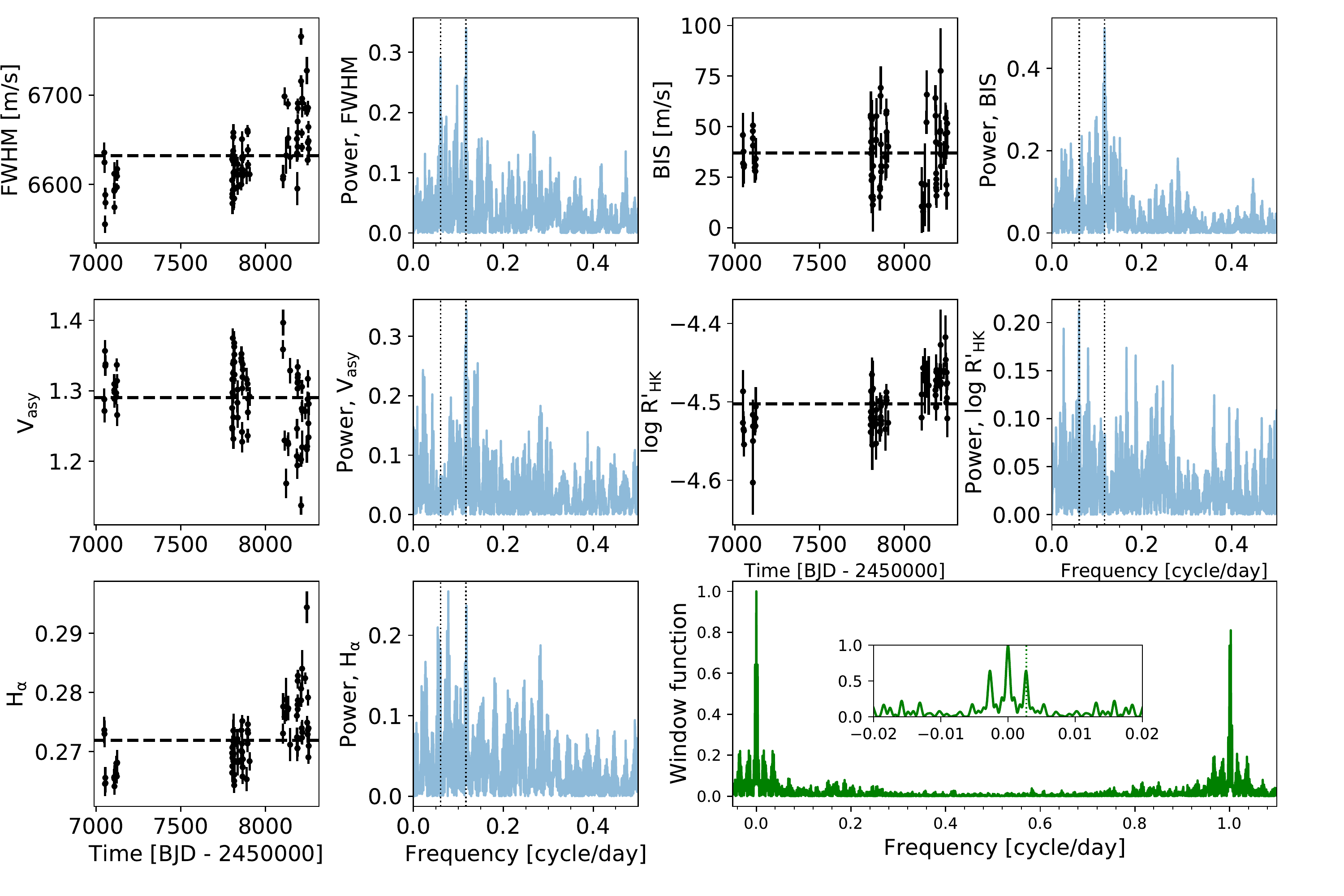}
      \caption{Time series of the FWHM, BIS and V$_{\rm asy}$ CCF line diagnostics, and the log(R$\ensuremath{'}_{\rm HK}$) and H$_{\rm \alpha}$ activity indexes, shown next to their corresponding GLS periodograms (after removing the trends in the FWHM, log(R$\ensuremath{'}_{\rm HK}$) and H$_{\rm \alpha}$ data as described in the text). The dashed horizontal lines mark the mean values, to highlight the increasing trend visible in the FWHM, log(R$\ensuremath{'}_{\rm HK}$), and H$_{\rm \alpha}$ data. Vertical dotted lines in the periodograms mark the locations of the stellar rotation period, as derived from K2 photometry, and its first harmonic. The panel in the bottom right corner shows the window function of the data. A zoomed view of the low-frequency part of the spectrum is shown in the inset plot, and a vertical dotted line marks the 1-year orbital frequency of the Earth.}
         \label{Fig:actindserieperiod}
\end{figure*}

We characterized the activity of K2-36 during the time span of the HARPS-N observations by analysing a set of standard spectroscopic indicators.
We considered the full width at half maximum (FWHM) of the CCF, the bisector inverse slope (BIS) and V$_{\rm asy}$ indicators, that quantify the CCF line asymmetry (V$_{\rm asy}$ is defined by \citealt{figueira13}), the chromospheric activity index log R$\ensuremath{'}_{\rm HK}$, and the activity indicator based on the H$_{\rm \alpha}$ line (derived following the method described in \citealt{gomesdasilva11}).
The time series of all the indicators are listed in Table \ref{Table:actind} and shown in Fig. \ref{Fig:actindserieperiod}. Extending to HARPS-N the results by \cite{santerne15} valid for the HARPS spectrograph, the uncertainties of the FWHM and BIS are fixed to 2$\sigma_{\rm RV}$, where $\sigma_{\rm RV}$ are the RV internal errors. 
Positive trends are clearly visible for the FWHM, log R$\ensuremath{'}_{\rm HK}$, and H$_{\rm \alpha}$ index, pointing out that the level of activity of K2-36 increased during the time span of our observations. These trends could be part of a long-term activity cycle, but our time span of $\sim$3.5 years is not extended enough to confirm this. We note that the dispersion in the FWHM, BIS, V$_{\rm asy}$ and H$_{\rm \alpha}$ data increased with time, as a consequence of the increasing levels of activity. 
We calculated the frequency spectrum of these datasets using the Generalised Lomb-Scargle (GLS) algorithm \citep{zech09}, after correcting the trends in the FWHM, log $R\ensuremath{'}_{\rm HK}$, and H$_{\rm \alpha}$ data by subtracting from each seasonal data chunk the corresponding average value. The periodograms are shown in Fig. \ref{Fig:actindserieperiod}. For the FWHM, BIS, and V$_{\rm asy}$ we find the highest peak at 8.6 days; for log $R\ensuremath{'}_{\rm HK}$ the peak with the highest power occurs at 16.7 days, and for the H$_{\rm \alpha}$ index the main peak is at 12.9 days, even though peaks of slightly lower power occur at $P_{\rm rot}$ and its first harmonic. In conclusion, these indicators appear modulated over $P_{\rm rot}$ or $P_{\rm rot}$/2. We note that a strong correlation exists between log $R\ensuremath{'}_{\rm HK}$ and H$_{\rm \alpha}$ ($\rho_{\rm Spear}$=0.8). 

We fitted all the indicators with a quasi-periodic GP model, as explained in Sect. \ref{sect:lcanalisi}. Since they have been extracted from the same spectra used to measure the RVs, the outcomes of such analysis are important to eventually set up the analysis of the RVs, in order to properly remove the stellar activity contribution. Results of the GP regression are listed in Table \ref{Tab:actindgp}. 

\begin{table}
  \caption[]{Best-fit values obtained from modelling the transit light curve of the K2-36 planets. Data from literature are shown for comparison. }
         \label{Table:planetfittransit}
         \centering
         \small
   \begin{tabular}{lll}
            \hline
            \noalign{\smallskip}
            Parameter &  Best-fit value & Reference \\
            \noalign{\smallskip}
            \hline
            \noalign{\smallskip}
            $P_{\rm p, b}$ [days] & 1.422614$\pm$0.000038 & (1) \\
            & 1.42266$\pm$0.00005 & (2)\\
            \noalign{\smallskip}
            & 1.422619$\pm$0.000039 & (3)\\
            \noalign{\smallskip}
            $T_{\rm 0, b}$ [BJD-2\,454\,833] & 1977.8916$\pm$0.0013  & (1)\\
            & 1977.8914$\pm$0.0013 & (3)\\
            \noalign{\smallskip}
            $T_{\rm dur, b}$ [days] & 0.05157$^{+0.0046}_{-0.0034}$  & (1)\\
            \noalign{\smallskip}
            $P_{\rm p, c}$ [days] & 5.340888$\pm$0.000086 & (1) \\
            \noalign{\smallskip}
            & 5.34059$\pm$0.00010 & (2)\\
            \noalign{\smallskip}
            & 5.340883$^{+0.000088}_{-0.000089}$ & (3)\\
            \noalign{\smallskip}
            $T_{\rm 0, c}$ [BJD-2\,454\,833] & 1979.84001$\pm$0.00071 & (1)\\
            \noalign{\smallskip}
            & 1979.84015$^{+0.00072}_{-0.00073}$ & (3)\\
            \noalign{\smallskip}
            $T_{\rm dur, c}$ [days] &  0.0564$^{+0.0035}_{-0.0027}$  & (1) \\
            \noalign{\smallskip}
            $R_{\rm b}$/R$_{\rm *}$ & 0.01828$^{+0.00072}_{-0.00099}$ & (1)\\
            \noalign{\smallskip}
            & 0.0180$^{+0.0026}_{-0.0013}$ & (3) \\ 
            \noalign{\smallskip}
            $R_{\rm c}$/R$_{\rm *}$ & 0.04119$^{+0.0043}_{-0.0029}$ & (1)\\
            \noalign{\smallskip}
            & 0.04$^{+0.17}_{-0.01}$ & (3) \\ 
            \noalign{\smallskip} 
            $R_{\rm b}$ (R$_{\rm \oplus}$) & 1.43$\pm$0.08 & (1) \\ 
            \noalign{\smallskip}
            & 1.32$\pm$0.09 & (2) \\
            \noalign{\smallskip}
            & 1.47$^{+0.22}_{-0.11}$ & (3) \\
            \noalign{\smallskip} 
            $R_{\rm c}$ (R$_{\rm \oplus}$) & 3.2$\pm$0.3 & (1) \\ 
            \noalign{\smallskip}
            & 2.80$^{+0.43}_{-0.31}$ & (2) \\
            \noalign{\smallskip}
            & 4.0$^{+14.0}_{-1.0}$ & (3) \\
            \noalign{\smallskip} 
            $i_{\rm b}$ [deg] & 84.45$^{+0.78}_{-0.48}$ & (1) \\ 
            \noalign{\smallskip}
            & 86.3$^{+2.7}_{-6.2}$ & (3) \\ 
            \noalign{\smallskip} 
            $i_{\rm c}$ [deg] & 86.917$^{+0.066}_{-0.056}$ & (1) \\ 
            \noalign{\smallskip}
            & 85.5$^{+3.9}_{-3.3}$ & (3) \\ 
            \noalign{\smallskip} 
            $b_{\rm b}$  & 0.66$^{+0.06}_{-0.09}$ & (1) \\ 
            \noalign{\smallskip} 
            $b_{\rm c}$  & 0.89$^{+0.01}_{-0.02}$ & (1) \\  
            \noalign{\smallskip} 
            $a_{\rm b}$ [R$_{\rm \star}$] & 6.63$\pm$0.11 & (1) \\ 
            \noalign{\smallskip}
            $a_{\rm b}$ [AU] & 0.0223$\pm$0.0004 & (1) \\
            \noalign{\smallskip} 
            $a_{\rm c}$ [R$_{\rm \star}$] & 16.01$^{+0.26}_{-0.27}$ & (1) \\ 
            \noalign{\smallskip}
            $a_{\rm c}$ [AU] & 0.054$\pm$0.001 & (1) \\ 
            \noalign{\smallskip}
            $T_{\rm eq, b}$ [K] & 1224$\pm$13 & (1a) \\ 
            \noalign{\smallskip} 
            $T_{\rm eq, c}$ [K] & 788$\pm$9 & (1a)\\  
            \noalign{\smallskip}
            Insolation $S_{\rm b} [S_{\rm \oplus}$] & 529$\pm23$ & (1) \\  
            \noalign{\smallskip}
            Insolation $S_{\rm c} [S_{\rm \oplus}$] & 90$\pm4$ & (1) \\
            \noalign{\smallskip}
            \hline
     \end{tabular}    
     \tablefoot{
     (1) This work. (1a) This work, assuming the Bond albedo $A_{\rm b}$=0.3 (2) \cite{sinukoff16}. (3) \cite{mayo18}
}
\end{table}

\begin{table*}
  \caption{GP hyper-parameters of a quasi-periodic model applied to the K2 light curve (6-hr bins), and to the asymmetry and activity index time series derived from HARPS-N spectra. Uncertainties are given as the $16^{\rm th}$ and $84^{\rm th}$ percentiles of the posterior distributions.}
         \label{Tab:actindgp}
         \small
   \begin{tabular}{llcccccc}
            \hline
            \noalign{\smallskip}
            Jump parameter  & Prior & K2 light curve & BIS span & FWHM\tablefootmark{a} & V$_{\rm asy}$\tablefootmark{a} & log $R\ensuremath{'}_{\rm HK}$\tablefootmark{a} & H$_{\rm \alpha}$\tablefootmark{a} \\
            \noalign{\smallskip}
            \hline
            \noalign{\smallskip}
            $h$  & light curve: $\mathcal{U}$(0,0.5) & 0.31$_{\rm -0.04}^{\rm +0.06}$ & 15$_{\rm -3}^{\rm +4}$ & 40$_{\rm -6}^{\rm +8}$ & 0.7$\pm$0.1 & 0.025$_{\rm -0.006}^{\rm +0.009}$ & 0.11$\pm$0.02 \\
                & BIS-$\braket{BIS}$: $\mathcal{U}$(0,100) [\ms] \\
                 & FWHM-$\braket{FWHM}$: $\mathcal{U}$(0,200) [\ms] & & & & \\
                 & V$_{\rm asy}$: $\mathcal{U}$(0,2) [\ms] & & & & \\
                 & $\log R\ensuremath{'}_{\rm HK}$ - $\braket{\log R\ensuremath{'}_{\rm HK}}$ : $\mathcal{U}$(0,1) [dex] & & & & \\
                 & H$_{\rm \alpha}: \mathcal{U}$(0,0.5) & & & & \\
                             \noalign{\smallskip}
            $\lambda$  & $\mathcal{U}$(0,1500) [days]& 106$_{\rm -11}^{\rm +14}$ & 144$_{\rm -55}^{\rm +100}$ & 79$_{\rm -24}^{\rm +34}$ & 161$_{\rm -54}^{\rm +80}$ &  270$_{\rm -111}^{\rm +175}$ & 1308$_{\rm -251}^{\rm +142}$ \\
                        \noalign{\smallskip}
            $w$ & $\mathcal{U}$(0,1) & 0.987$_{\rm -0.02}^{\rm +0.009}$ & 0.6$\pm$0.2 & 0.41$_{\rm -0.06}^{\rm +0.08}$ & 0.87$\pm$0.1 & 0.8$_{\rm -0.2}^{\rm +0.1}$ & 0.96$_{\rm -0.06}^{\rm +0.03}$ \\
                        \noalign{\smallskip}
            $\theta$ & $\mathcal{U}$(0,20) [days] & 16.9$\pm$0.2 & 8.56$_{\rm -0.04}^{\rm +0.07}$ & 17.2$\pm$0.1 & 17.24$_{\rm -0.53}^{\rm +0.07}$ & 16.7$_{\rm -3.2}^{\rm +0.3}$ & 16.90$\pm$0.01 \\
            \noalign{\smallskip}
            \hline
     \end{tabular}  
            \tablefoottext{a}{Time series not corrected for the long-term trend.}
\end{table*}

\onllongtab{
\begin{longtable}{ccccccccc}
\caption{Spectroscopic activity/CCF asymmetry indicators for K2-36. Data are extracted from the 81 spectra used in this work (the full dataset is available on-line via CDS).}
\label{Table:actind}\\
\hline  \hline
\textbf{Time }& \textbf{FWHM} & \textbf{BIS} & \textbf{log(R$\ensuremath{'}_{\rm HK}$)} & \textbf{$\sigma_{\rm log(R\ensuremath{'}_{\rm HK})}$} & \textbf{H$_{\rm \alpha}$} & \textbf{$\sigma_{\rm H_{\rm \alpha}}$} & \textbf{V$_{\rm asy}$} \tablefootmark{a} & \textbf{$\sigma_{\rm V_{\rm asy}}$} \\
(BJD$_{\rm TDB}$-2\,400\.000) & ($\ms$) & ($\ms$) &  &  &  & \\
\hline
\noalign{\smallskip}
\hline
\endfirsthead
\caption{Continued.} \\
\hline\hline
\noalign{\smallskip}
\textbf{Time }& \textbf{FWHM} & \textbf{BIS} & \textbf{log(R$\ensuremath{'}_{\rm HK}$)} & \textbf{$\sigma_{\rm log(R\ensuremath{'}_{\rm HK})}$} & \textbf{H$_{\rm \alpha}$} & \textbf{$\sigma_{\rm H_{\rm \alpha}}$} & \textbf{V$_{\rm asy}$}\footnotemark{a} & \textbf{$\sigma_{\rm V_{\rm asy}}$}  \\
(BJD$_{\rm TDB}$-2\,400\.000) & ($\ms$) & ($\ms$) &  &  & & \\
\noalign{\smallskip}
\hline
\noalign{\smallskip}
\endhead
\hline
\endfoot
\hline
\endlastfoot
\noalign{\smallskip}
57047.651272 & 6635.69 & 45.79 & -4.5267 & 0.0270 & 0.27365 & 0.00221 & 1.28786 & 0.01556 \\
57048.697294 & 6624.72 & 31.80 & -4.4866 & 0.0270 & 0.27298 & 0.00229 & 1.27135 & 0.01631 \\
57051.659658 & 6555.43 & 37.65 & -4.5348 & 0.0229 & 0.26454 & 0.00209 & 1.35659 & 0.01546 \\
... & ... & ... & ... & ... & ... & ... & ... & ...\\
\end{longtable}
}


\section{RV analysis}
\label{sect:rvanalysis}
Predicting the amplitude of the RV variations due to the stellar contribution based on the photometric variability is a complex task, especially when photometric and spectroscopic data are not collected during the same seasons. Following \cite{mann18}, a guess for the RV variability could be estimated as $\sim$ $\sigma_{\rm phot}\cdot v \sin i_{\rm \star}$, where $\sigma_{\rm phot}$ is the scatter observed in the K2 light curve. The expected RV variability for K2-36 due to stellar activity is therefore $\sim$2-5 \ms. 
According to the calibration for K-dwarfs derived by \cite{santos2000}, the expected radial velocity scatter for K2-36 is $\sim$9 \ms.
We anticipate here that the signal due to the stellar activity largely dominates the observed RVs variations, and it is 5-8 times larger than the semi-amplitudes of the planetary signals, which are of the order of the average RV internal errors ($\sim$2-3 \ms). 
When dealing with such an active star it is interesting to compare the results obtained from RV extracted with two different procedures, that in principle could be sensitive to stellar activity in different ways for this specific case.   
Therefore, we have calculated the systemic RVs through the CCF-based recipe of the on-line DRS, using a mask suitable for stars with spectral type K5V, and relative RVs with the TERRA pipeline \citep{anglada12}, measured against a high S/N template spectrum. We corrected the TERRA RVs for the perspective acceleration using parallax and proper motion from \textit{Gaia} DR2.
The complete list of RVs is presented in Table \ref{Table:radvel}.
We summarize general properties of the two RV datasets in Table \ref{Tab:rvproperties}, and show the time series in Fig. \ref{Fig:rvtimeperiod}. One common feature is the increase of the scatter observed in the last two seasons with respect to the first. This is a consequence of the increasing stellar activity discussed in Sect. \ref{sect:actindex}. The scatter over the time span is similar for all the methods, and the internal errors calculated with TERRA are on average lower than those of the DRS.   

After a preliminary analysis of the RVs, in this Section we present, and compare, the results of different global models (combined fit of planet orbital equations and stellar signals). We used the results from the light curve analysis to fix the priors of the orbital periods and time of inferior conjunction for K2-36\,b and K2-36\,c. We assume independent Keplerian motions (i.e. ignoring mutual planetary perturbations, which would be significantly below 1 mm s$^{\rm -1}$) in all the cases and, especially considering the short orbital period of K2-36\,b, we did not bin the data on a nightly basis to make use of all the available information in each single data point. 

\subsection{Frequency content and correlation with activity indicators}
We scrutinized the frequency content in the RV datasets with the GLS algorithm by fitting a single sinusoid to the data. Periodograms are shown in Fig. \ref{Fig:rvtimeperiod}. All look very similar, with the main peak clearly located at $\sim$8.5 days, that is the first harmonic of $P_{\rm rot}$, and semi-amplitudes of $\sim$16 \ms, which are equal to the measured RMS.  
This analysis shows that signals due to stellar activity dominate the RVs. The orbital frequency $f\sim$0.7 d$^{\rm -1}$ associated with K2-36\,b appears with much less power, while the Doppler signal of K2-36\,c is missing. We analysed the residuals of the RVs with a GLS periodogram (using DRS data only because of the similarities with the periodogram of the TERRA RVs), after subtracting the best-fit sinusoid, and the periodogram shows the main peak at $P_{\rm rot}$. An additional pre-whitening results in a peak at frequency $f\sim0.17$ d$^{\rm -1}$, which likely corresponds to the second harmonic of $P_{\rm rot}$. This analysis points out that the stellar activity component could be mitigated by a sum of at least three sinusoids with periods $P_{\rm rot}$, $P_{\rm rot}$/2 and $P_{\rm rot}$/3. This simple pre-whitening procedure does not allow for the identification of significant power at the orbital planetary frequencies, pointing out that the detection of the two planets using only RVs, without the help from transits and an effective treatment of the stellar activity contribution, is very complicated. 
We have folded the RVs of each season taken separately at the period P=8.5 days, and we observe a change in the phase of the signal. This suggests that the stellar activity signal has been evolving on a timescale less than one year, and that a quasi-periodic fit of the stellar component could be more appropriate than a simple combination of sinusoids.    

\begin{figure}
   \centering
   \includegraphics[width=\hsize]{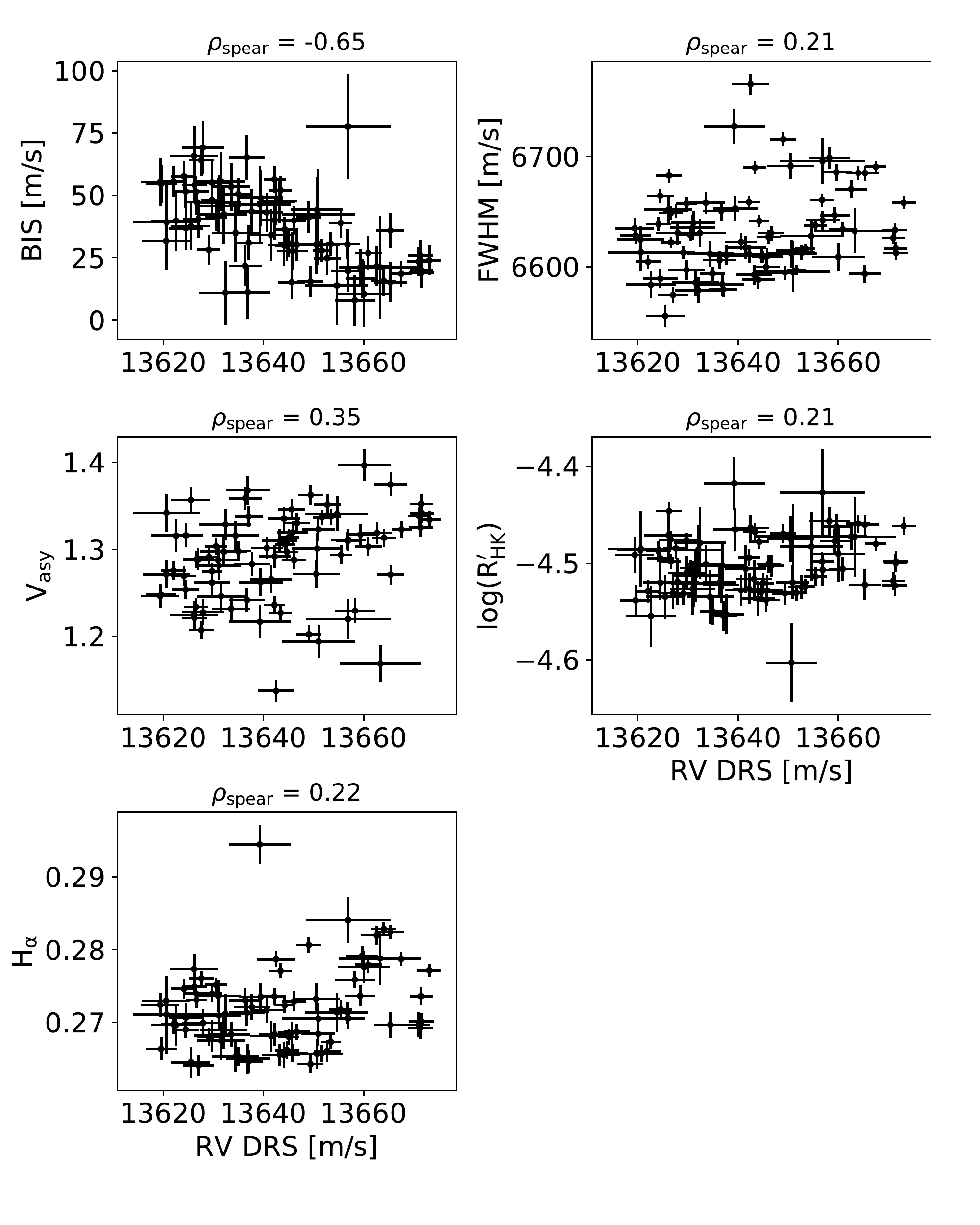}
      \caption{Correlations between radial velocities (from DRS), the CCF line profile indicators, and activity diagnostics.}
         \label{Fig:rvactindcorr}
\end{figure}

We checked the correlation between the RVs and the CCF line asymmetry and activity indicators discussed in Sect. \ref{sect:actindex} by calculating the Spearman's rank correlation coefficients. Results are shown in Fig. \ref{Fig:rvactindcorr}, where we used the RVs from DRS. Significant anti-correlation exists just between the DRS RVs and BIS\footnote{Since the BIS span indicator is extracted with the DRS pipeline, we limit this analysis to the DRS RVs only.} ($\rho_{\rm Spear}$ = -0.65), which is not surprising since they have similar GLS periodograms. This anti-correlation is expected when the observed RV variations are due mainly to the presence of dark spots on the stellar photosphere, as shown by the light curve modulation \citep{boisse11}. 

\subsection{Joint modelling of planetary and stellar signals (1): sum of sinusoids}
\label{subsec:model1}
Based on the results of the GLS analysis, first we tested a model (hereafter model 1) with the activity component described by a sum of three sinusoids, each with periods sampled around $P_{\rm rot}$, $P_{\rm rot}$/2 and $P_{\rm rot}$/3, using the priors indicated in Table 7. 
We modelled the planetary signals as circular orbits. In total, 17 free parameters were used, including the uncorrelated jitter and offset, that for the DRS RVs corresponds to the systemic velocity.
Table \ref{Tab:mcmcfitrvsinusoid} shows the results from the analysis of the DRS and TERRA datasets. The semi-amplitude $K_{\rm b}$ is fitted with a relative error of $\sim50\%$, while that of the signal induced by K2-36\,c is nearly twice $K_{\rm b}$. We note that the required uncorrelated jitter $\sigma_{\rm jit}$ is more than twice the typical internal error $\sigma_{\rm RV}\sim$2-3 \ms.

\onllongtab{
\begin{longtable}{c|cc|cc}
\caption{Radial velocities of K2-36 extracted from the 81 HARPS-N spectra analyzed in this work (the full dataset is available on-line via CDS).}
\label{Table:radvel}\\
\hline  \hline
& \multicolumn{2}{c|}{DRS} & \multicolumn{2}{c}{TERRA} \\ 
\hline
\textbf{Time }& \textbf{Radial velocity} & \textbf{$\sigma_{\rm RV}$} & \textbf{Radial velocity} & \textbf{$\sigma_{\rm RV}$} \\ 
(BJD$_{\rm TDB}$-2\,400\.000) & ($\ms$) & ($\ms$) & ($\ms$) & ($\ms$) \\ 
\hline
\noalign{\smallskip}
\hline
\endfirsthead
\caption{Continued.} \\
\hline\hline
\noalign{\smallskip}
& \multicolumn{2}{c|}{DRS} & \multicolumn{2}{c}{TERRA} \\ 
\hline
\textbf{Time }& \textbf{Radial velocity} & \textbf{$\sigma_{\rm RV}$} & \textbf{Radial velocity} & \textbf{$\sigma_{\rm RV}$} \\ 
(BJD$_{\rm TDB}$-2\,400\.000) & ($\ms$) & ($\ms$) & ($\ms$) & ($\ms$) \\ 
\noalign{\smallskip}
\hline
\noalign{\smallskip}
\endhead
\hline
\endfoot
\hline
\endlastfoot
\noalign{\smallskip}
     57047.651272 & 13630.93 & 4.44 &  -9.07 & 3.55 \\ 
   57048.697294 & 13620.53 & 4.76 & -21.54 & 3.20 \\ 
   ... & ... & ... & .. & .. \\
\end{longtable}
}

\begin{table*}
  \caption{Properties of the two RV datasets analysed in this work.}
         \label{Tab:rvproperties}
         \small
         \centering
   \begin{tabular}{c c c c c c}
            \hline
            \noalign{\smallskip}
            Pipeline &  RMS (all data) & RMS (1$^{\rm st}$ seas.) & RMS (2$^{\rm nd}$ seas.) & RMS (3$^{\rm rd}$ seas.) & Median internal error $\sigma_{\rm RV}$ \\
                       &  [\ms] &  [\ms] &  [\ms] &  [\ms] & [\ms] \\
            \noalign{\smallskip}
            \hline
            \noalign{\smallskip}
            DRS & 14.9 & 9.8 & 15.6 & 15.5 & 3.1 \\
            TERRA & 14.7 & 9.4 & 14.8 & 16.2 & 2.3 \\
            \noalign{\smallskip}
            \hline
     \end{tabular}  
\end{table*}

\begin{table*}
  \caption{Results of the MC analysis performed on the RVs extracted with the DRS and TERRA recipes using a sum of three sinusoids to model the stellar activity component. Uncertainties are given as the $16^{\rm th}$ and $84^{\rm th}$ percentiles of the posterior distributions.}
         \label{Tab:mcmcfitrvsinusoid}
         \small
         \centering
   \begin{tabular}{l c c c}
            \hline
            \noalign{\smallskip}
            Jump parameter &  Prior &  \multicolumn{2}{c}{Best-fit value} \\
            & & DRS & TERRA \\
            \noalign{\smallskip}
            \hline
            \noalign{\smallskip}
            $\gamma$ [\ms] & $\mathcal{U}$(13550,13750) [DRS] & 13643.6$^{+0.8}_{-0.9}$ & -0.09$^{+0.84}_{-0.86}$ \\
            & $\mathcal{U}$(-100,100) [TERRA] & & \\
            \noalign{\smallskip}
            $\sigma_{\rm jit}$ [\ms] & $\mathcal{U}$(0,20) & 6.4$^{\rm +0.8}_{\rm -0.7}$ & 6.8$^{\rm +0.7}_{\rm -0.6}$ \\
            \noalign{\smallskip}
            \textbf{Stellar activity signal} & & \\
            \noalign{\smallskip}
            $P_{\rm rot}$ [days] & $\mathcal{U}$(15,18) & 16.65$\pm$0.02 & 16.68$\pm$0.02 \\
            \noalign{\smallskip}
            $P_{\rm rot}$/2 [days] & $\mathcal{U}$(7,9.5) & 8.499$^{\rm +0.006}_{\rm -0.005}$ & 8.507$\pm$0.005 \\
            \noalign{\smallskip}
            $P_{\rm rot}$/3 [days] & $\mathcal{U}$(4.5,6.5) & 5.754$\pm$0.003 & 5.755$^{\rm +0.004}_{\rm -0.138}$ \\
            \noalign{\smallskip}
            $K_{\rm act,P=P_{\rm rot}}$ [\ms] & $\mathcal{U}$(0,30) & 7.8$\pm$1.2 & 7.9$\pm$1.2 \\
            \noalign{\smallskip}
            $T_{\rm 0,act,P=P_{\rm rot}}$ [BJD-2\,450\,000] & $\mathcal{U}$(7800,7820) & 7811.5$\pm$0,4 & 7811.9$^{+0.5}_{-0.4}$ \\ 
            \noalign{\smallskip}
            $K_{\rm act,P=P_{\rm rot}/2}$ [\ms] & $\mathcal{U}$(0,30) & 14.6$\pm$1.1 & 14.5$\pm$1.1\\
            \noalign{\smallskip}
            $T_{\rm 0,act,P=P_{\rm rot}/2}$ [BJD-2\,450\,000] & $\mathcal{U}$(7800,7810) & 7809.4$\pm$0.2 & 7809.1$\pm$0.2 \\ 
            \noalign{\smallskip}
            $K_{\rm act,P=P_{\rm rot}/3}$ [\ms] & $\mathcal{U}$(0,20) & 7.4$\pm$1.4 & 6.7$\pm$1.3\\
            \noalign{\smallskip}
            $T_{\rm 0,act,P=P_{\rm rot}/3}$ [BJD-2\,450\,000] & $\mathcal{U}$(7800,7808) & 7802.9$\pm$0.2 & 7802.8$^{\rm +0.3}_{\rm -0.2}$ \\ 
            \noalign{\smallskip}         
            \textbf{Planetary signals} & & \\
            \noalign{\smallskip}
            $K_{\rm b}$ [\ms] & $\mathcal{U}$(0,10) & 2.8$\pm$1.3 & 2.4$\pm$1.2 \\
            \noalign{\smallskip}
            $K_{\rm c}$ [\ms] & $\mathcal{U}$(0,10) & 5.0$\pm$1.4 & 4.9$\pm$1.4 \\
            \noalign{\smallskip}
            $m_{\rm b}$ [$\mearth$] & derived & 4.2$\pm$1.8 & 3.7$^{\rm +1.8}_{\rm -1.7}$ \\
            \noalign{\smallskip}
            $m_{\rm c}$ [$\mearth$] & derived & 11.7$^{\rm +3.0}_{\rm -3.4}$ & 11.5$\pm$3.2 \\
            \noalign{\smallskip}
            \textbf{Bayesian evidence ln($\mathcal{Z})_{\rm model\:1}$} & & -310.5$\pm$0.06 & -310.14$\pm$0.04\\
            \noalign{\smallskip}
            \hline
     \end{tabular}  
\end{table*}

\subsection{Joint modelling of planetary and stellar signals (2): de-trending using the BIS line asymmetry indicator}
Since a significant anti-correlation is observed between the DRS RVs and the BIS asymmetry indicator, we modelled the stellar signal contribution by including a linear term $c_{\rm BIS}\cdot$BIS in the model (hereafter model 2). The planetary orbits were fixed to the circular case. For the semi-amplitudes of the planetary signals we get estimates that are nearly twice and nearly half those derived for model 1, for planet b and c respectively: $K_{\rm b}$ = 4.8$^{+1.8}_{-1.7}$ \ms and $K_{\rm c}$ = 2.8$^{+1.7}_{-1.5}$ \ms. Since $\ln\left( \frac{\mathcal{Z}_{\rm model\:1}} {\mathcal{Z_{\rm model\:2}}} \right)\simeq$+10, which corresponds to a posterior odds ratio $\simeq2\times10^{\rm 4}$:1, this model is very significantly disfavoured, according to the conventional scale adopted for model selection (see, e.g., \citealt{feroz11}).

\subsection{Joint modelling of planetary and stellar signals (3): Gaussian process regression analysis}
We analysed the RV datasets with a GP regression using the quasi-periodic kernel (Eq. \ref{eq:1}) to model the correlated signal introduced by the stellar activity contribution. Since both signals modulated on $P_{\rm rot}$ and $P_{\rm rot}$/2 are present in the RV timeseries, we left the hyper-parameter $\theta$ free to explore uniformly the range between 0 and 20 days. In general, for the hyper-parameters $\theta$, $w$, and $\lambda$ we used the same priors as those in Table \ref{Tab:actindgp}. We included the parameter $\sigma_{\rm jit}$, which is summed in quadrature with $\sigma_{\rm RV}$ in Eq. \ref{eq:1}, to take into account other sources of uncorrelated noise not included in $\sigma_{\rm RV}(t)$.

We first considered the case of circular orbits, then we included the eccentricities $e_{\rm b}$ and $e_{\rm c}$ as free parameters.
The results of the analyses are shown in Table \ref{Table:percentilesgprv}. Concerning the circular model, looking at the best-fit values of the hyper-parameters, $\theta$ is determined with high precision and its value corresponds to $P_{\rm rot}$, and the evolutionary timescale of the active regions $\lambda$ is $\sim$5-8 times $P_{\rm rot}$, depending on the RV extraction algorithm. This is in very good agreement with the results of the analysis of the light curve and the FWHM activity diagnostic, while there is evidence for a slightly longer evolutionary timescale from the $\log$ $R\ensuremath{'}_{\rm HK}$ activity index. The Doppler semi-amplitude $K_{\rm b}$ is in agreement within the errors with that obtained from model 1, both for DRS ($K_{\rm b}$=2.1$\pm$0.9 \ms) and TERRA ($K_{\rm b}$=2.6$\pm$0.7 \ms), but more precise.  
The mass of K2-36\,b is measured with a significance of $\sim2.5\sigma$ (DRS, $m_{\rm b}=3.2^{\rm +1.4}_{\rm -1.3}$ $\mearth$) and $\sim$3.6$\sigma$ (TERRA, $m_{\rm b}=3.9\pm1.1$ $\mearth$), and there is agreement within 1$\sigma$ between the two RV extraction pipelines. As for K2-36\,c, the mass is measured at best with a significance of $\sim$3.5$\sigma$ (TERRA, $m_{\rm c}=7.8\pm2.3$ $\mearth$), and all the values are in agreement within 1$\sigma$. When compared to model 1, the semi-amplitude of the K2-36\,c signal is lower for both DRS and TERRA, and more precise. 

Our analysis does not constrain the eccentricities of the two planets. In fact, for K2-36\,b we get $e_{\rm b,\:DRS}=0.41^{+0.33}_{-0.30}$ (68.3$^{\rm th}$ percentile=0.57) and $e_{\rm b,\:TERRA}=0.51^{+0.26}_{-0.37}$ (68.3$^{\rm th}$ percentile=0.65), while for K2-36\,c we get $e_{\rm c,\: DRS}=0.18^{+0.23}_{-0.12}$ (68.3$^{\rm th}$ percentile=0.27) and $e_{\rm c,\:TERRA}=0.14^{+0.17}_{-0.10}$ (68.3$^{\rm th}$ percentile=0.21). For both planets, the eccentricity differs from zero with a significance less than 1.5$\sigma$, and the peak of the posterior distribution of $e_{\rm c}$ occurs at zero. Therefore we conclude that our data do not allow us to constrain the eccentricities. Moreover, the eccentric model is statistically disfavoured with respect to the circular model for both the DRS ($\ln\mathcal{Z}_{\rm circular}-\ln\mathcal{Z}_{\rm ecc}$=+2) and TERRA ($\ln\mathcal{Z}_{\rm circular}-\ln\mathcal{Z}_{\rm ecc}$=+2.5) datasets.

\begin{figure}
   \centering
   \includegraphics[width=\hsize]{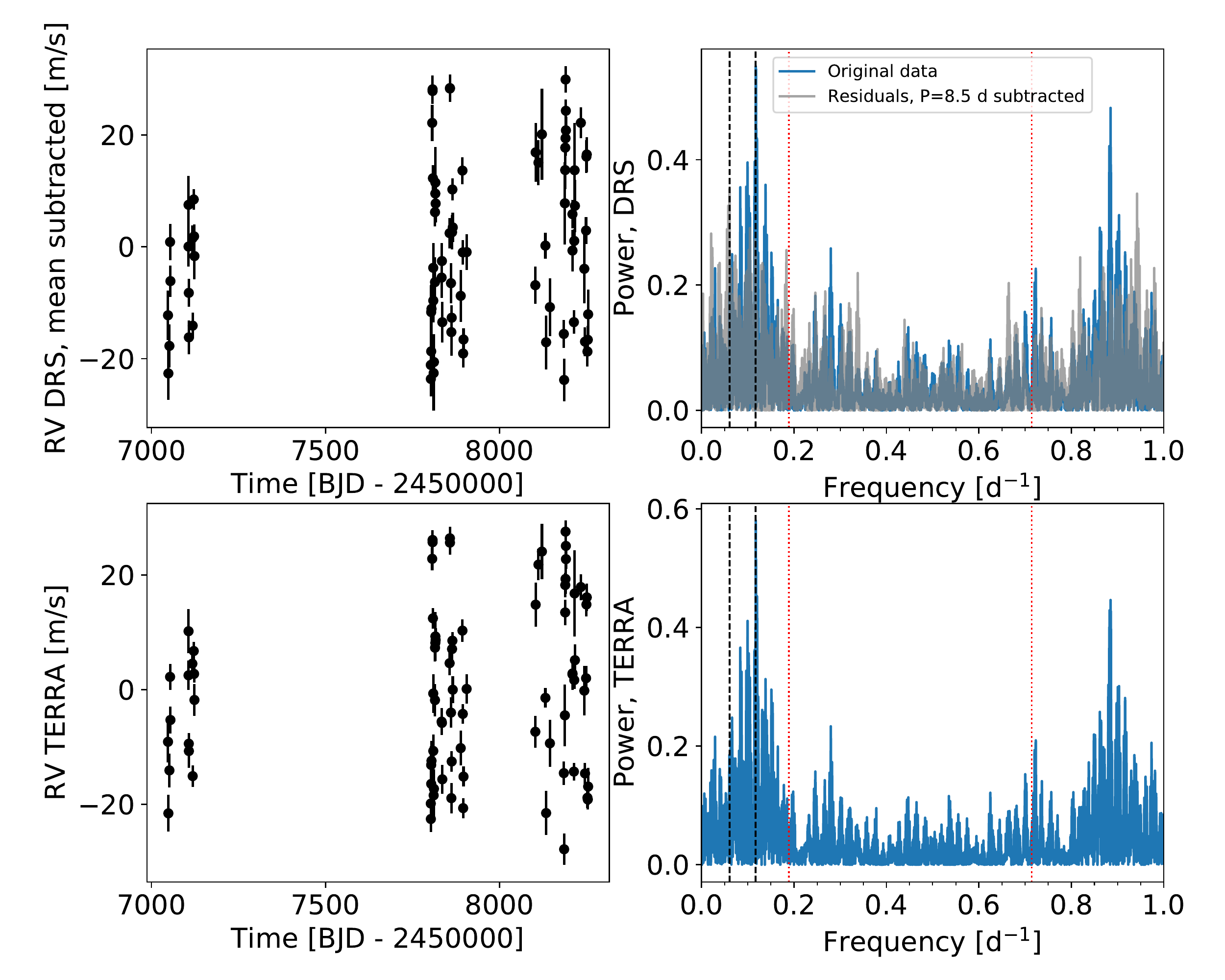}
      \caption{\textit{Left column}. Radial velocity time series extracted from HARPS-N spectra using the DRS (upper plot) and TERRA (lower plot) pipelines. \textit{Right column}. GLS periodograms of the RV time series (blue line). Vertical dashed lines mark the location of the highest peak at $\sim$8.5 days and the stellar rotation frequency, for all the datasets. Vertical dotted lines in red mark the orbital frequencies of the K2-36 planets. For the DRS dataset only, we show the periodogram of the residuals, after subtracting the signal with period of 8.5 days (gray line), with the main peak located at $P\sim$16.5 days, which corresponds to the stellar rotation period. The window function of the data (not shown) is the same as in Fig. \ref{Fig:actindserieperiod}.}
         \label{Fig:rvtimeperiod}
\end{figure}

\begin{figure}
   \centering
   \includegraphics[width=\hsize]{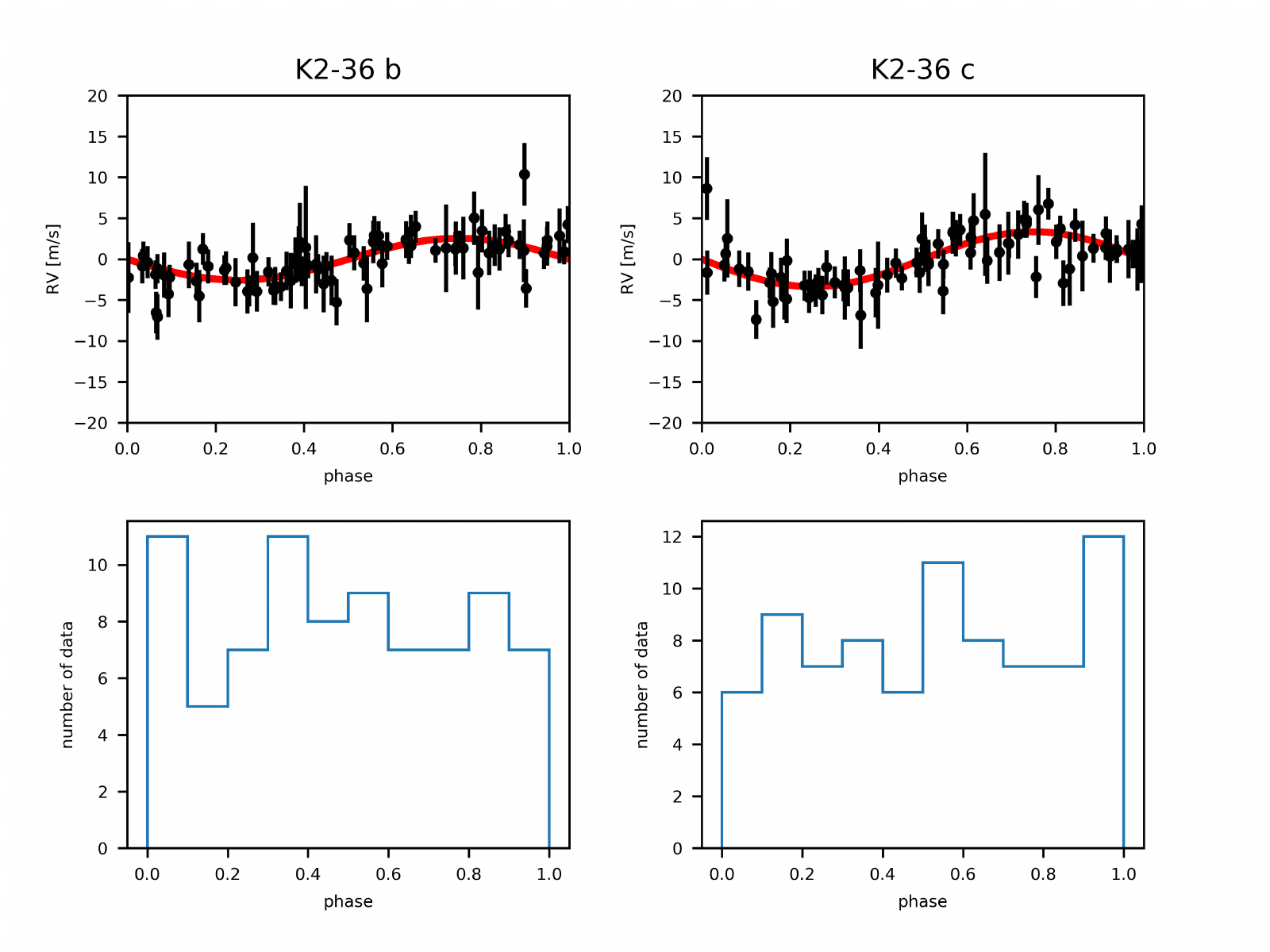}
      \caption{Doppler signals due to the K2-36 planets (TERRA radial velocity dataset, after removing the GP quasi-periodic stellar activity term), folded according to their transit ephemeris (phase=0 corresponds to the time of inferior conjunction). The histograms show the distribution of the RV measurements along the planetary orbits.}
         \label{Fig:rvplanetfold}
\end{figure}

\begin{figure}
   \centering
   \includegraphics[width=\hsize]{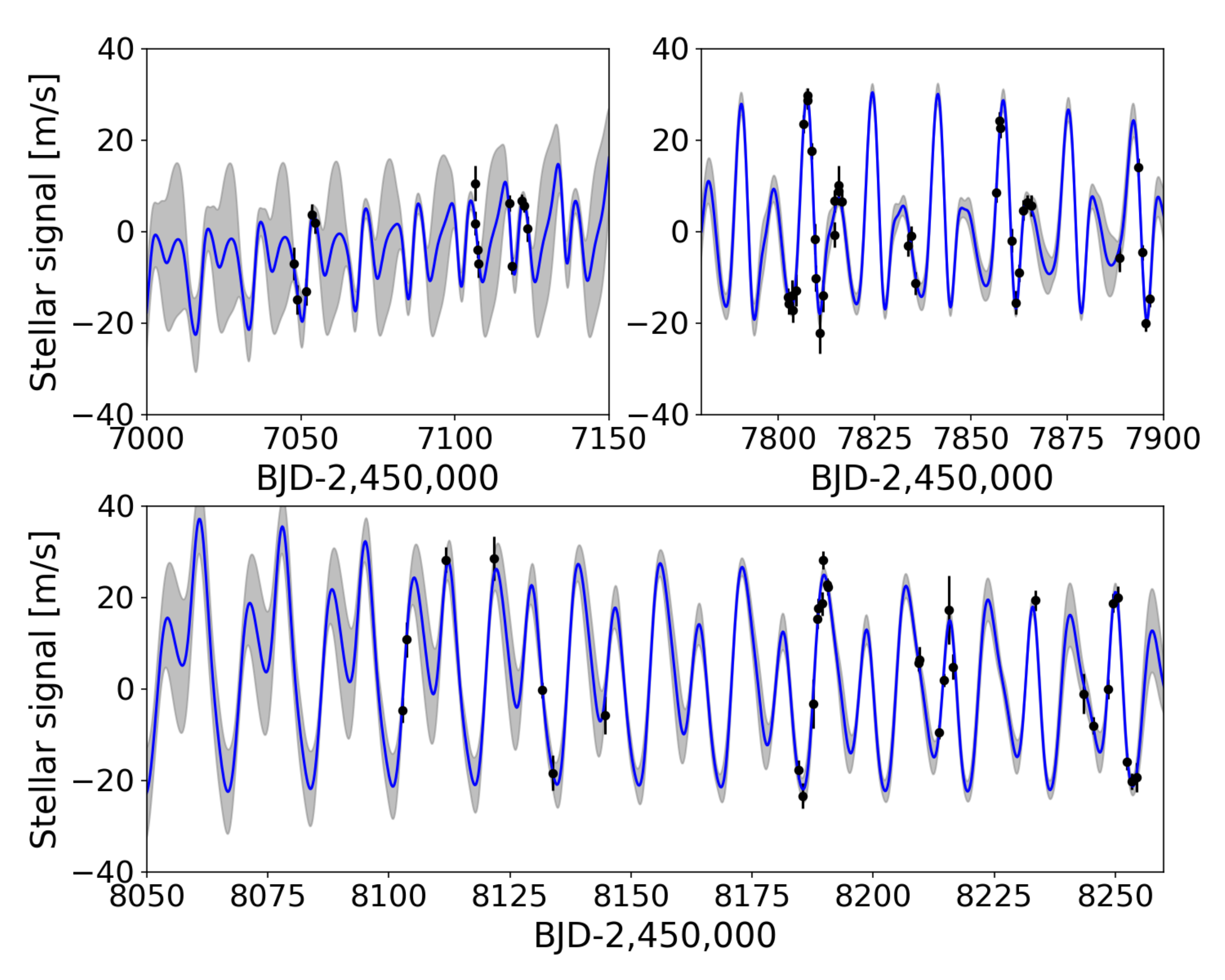}
      \caption{Stellar activity component in the radial velocities (TERRA dataset) as fitted using a GP quasi-periodic model (circular case). Each panel shows one of the three observing seasons. The blue line represents the best-fit solution, and the grey area the 1$\sigma$ confidence interval.}
         \label{Fig:rvstellarsignal}
\end{figure}

\begin{figure}
   \centering
   \includegraphics[width=\hsize]{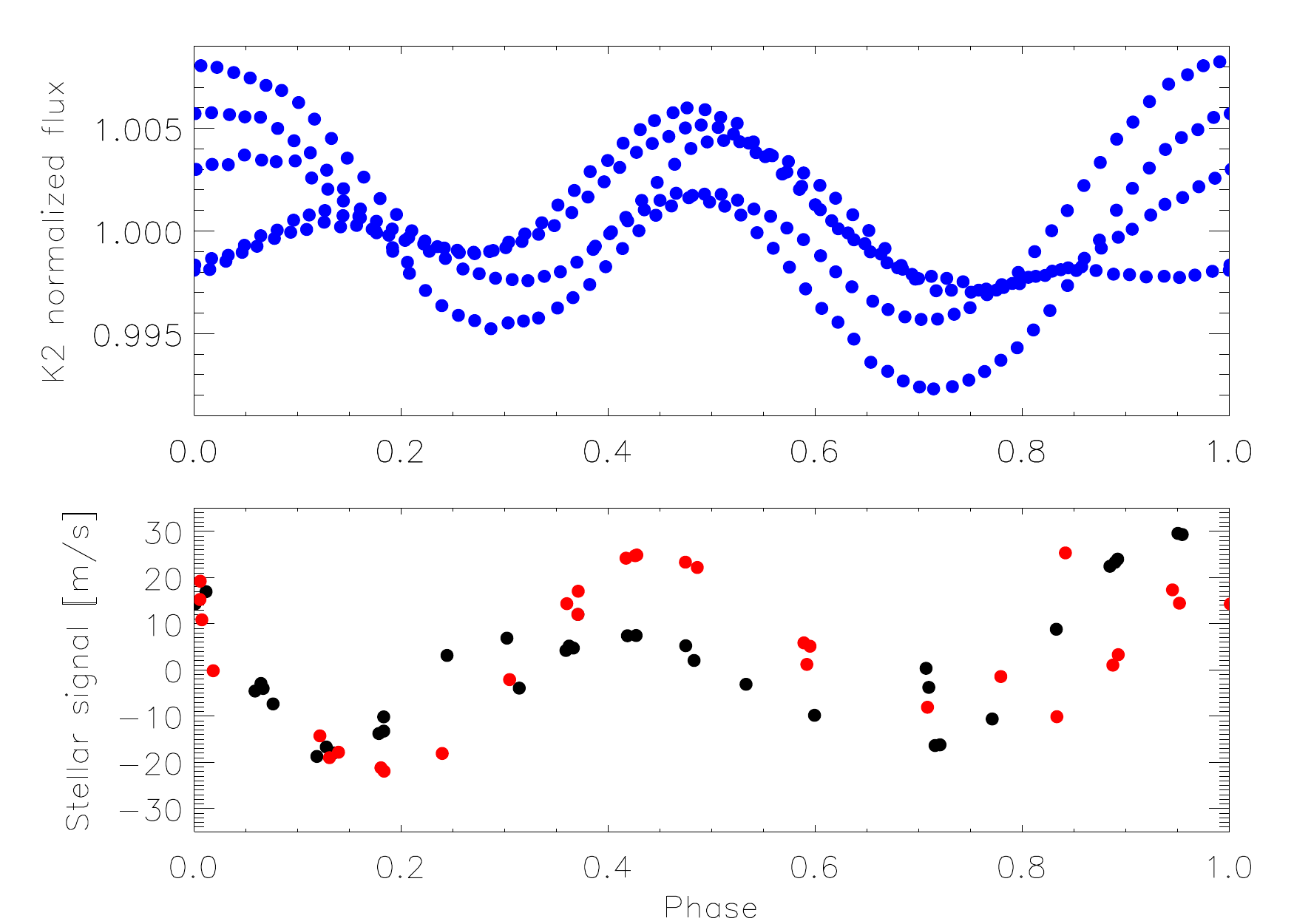}
      \caption{K2 light curve (upper plot) and stellar activity component in the radial velocities (lower plot; TERRA data for the last two seasons. DRS data show a similar behaviour), folded at the stellar rotation period. The reference epoch corresponding to phase=0 is the same for both datasets. Light curve goes from May, 30$^{\rm th}$ to Aug, 21$^{\rm th}$ 2014 (K2 campaign C1). Spectroscopic data are distinguished according the observing season: black dots are used for data collected in 2017, red dots for those of 2018.}
         \label{Fig:rvlcfold}
\end{figure}

Fig. \ref{Fig:rvplanetfold} shows the RV signals due to the K2-36 planets, and the distribution of the HARPS-N measurements along the planetary orbits. In both cases, the orbit has been covered uniformly. We show in Fig. \ref{Fig:rvstellarsignal} the stellar component present in the RVs as fit by the GP quasi-periodic model (TERRA dataset). 
We note that the GLS periodogram of the stellar signal only is very similar to that obtained for the original RV dataset (Fig. \ref{Fig:rvtimeperiod}), demonstrating that the quasi-periodic model gives a good representation of the activity component. Moreover, we folded the stellar component in the TERRA dataset, as fitted by the GP regression, at the stellar rotation period, and compared to the folded K2 light curve (Fig. \ref{Fig:rvlcfold}). Interestingly the RV data, collected in 2017 and 2018 and with a good phase coverage at $P_{\rm rot}$, have a similar phase pattern of the photometric data and appear only slightly shifted, despite more than 3 years separating the K2 observations from the HARPS-N observations.

\subsection{Adopted results}
We tested three different models to derive the main planetary parameters. According to their Bayesian evidences the GP quasi-periodic and circular model is strongly favoured over all the other models (in particular $\ln\left( \frac{\mathcal{Z}_{\rm model\:3}} {\mathcal{Z_{\rm model\:1}}} \right)\simeq$+14.5 for the TERRA dataset, and $\ln\left( \frac{\mathcal{Z}_{\rm model\:3}} {\mathcal{Z_{\rm model\:1}}} \right)\simeq$+9 for the DRS dataset), therefore we adopt this as the best representation for the K2-36 system based on our data.
Since the relative errors on the semi-amplitudes $K_{\rm b}$ and $K_{\rm c}$ are lower for the TERRA than for the DRS dataset, we adopt as our final solution the best-fit values obtained with the TERRA RVs. Based on that, K2-36\,c has a mass $\sim$2 times higher than that of K2-36\,b ($m_{\rm b}$=3.9$\pm$1.1 $\mearth$, $m_{\rm c}$=7.8$\pm$2.3 $\mearth$), but its bulk density is $\sim$19$\%$ that of the innermost planet ($\rho_{\rm b}$=7.2$^{+2.5}_{-2.1}$ g\,cm$^{-3}$, $\rho_{\rm c}$=1.3$^{+0.7}_{-0.5}$ g\,cm$^{-3}$).
By using the relation $S/S_{\rm \oplus}$ = ($L/L_{\rm \odot})\times$(AU/$a)^{\rm 2}$, we derived the insolation fluxes $S_{\rm b}$=529$\pm$23 \,S$_{\rm \oplus}$ and $S_{\rm c}$=90$\pm$4 \,S$_{\rm \oplus}$: planet b is $\sim$6 times more irradiated than K2-36\,c. The derived planet equilibrium temperatures are $T_{\rm eq, b}$=1224$\pm$13 K and $T_{\rm eq, c}$=788$\pm$9 K.

\subsection{Robustness of the derived planet masses}
We devised a test to assess how robust our results are. We simulated $N_{\rm sim}$=100 RV time series using the epochs of the HARPS-N spectra as time stamps, and assuming the TERRA dataset and the results of the GP global model to build the mock datasets. We added two planetary signals (circular orbits) to the quasi-periodic stellar activity signal, with semi-amplitudes $K_{\rm b}$=2.6 and $K_{\rm c}$=3.4 \ms. The orbital periods and epochs of inferior conjunction are those derived from K2 light curve. Each mock dataset has been obtained by randomly shifting each point of the exact solution within the error bar, assuming normal distributions centred to zero and with $\sigma=\sigma_{\rm RV}(t)$. We use this set of simulated data to test our ability to retrieve the injected signals. We analysed the mock RV time series within the same MC framework used to model the original dataset, and for each free parameter we saved the 16$^{\rm th}$, 50$^{\rm th}$ and 84$^{\rm th}$ percentiles. Then, we analysed the posterior distributions of the $N_{\rm sim}$ median values, and for the planetary Doppler semi-amplitudes we get $K_{\rm b}$=2.7$^{\rm +0.4}_{\rm -0.5}$ and $K_{\rm c}$=3.4$^{\rm +0.8}_{\rm -0.6}$ \ms (16$^{\rm th}$, 50$^{\rm th}$ and 84$^{\rm th}$ percentiles). Since these values are in very good agreement with those injected, that are equal to those we get form the GP analysis, this suggests that the best-fit values obtained for the original dataset are accurate. 


\subsection{Sensitivity of the planetary masses to an extended dataset}
Through a different set of simulations we have investigated how much the precision of the planet masses would have improved with an additional season of RV data. We simulated $N_{\rm sim}$=50 RV datasets by randomly drawing 40 additional epochs (different for each mock dataset) within a 180-day timespan, six months after the end of the third observing season with HARPS-N (for a total of 121 RV measurements per dataset, including the original data). We assumed a sampling characterized by one RV measurement per night, and adopted the GP quasi-periodic best-fit solution to represent the stellar activity term. We used the \texttt{sample$\_$conditional} module in the \texttt{GEORGE} package to draw samples from the predictive conditional distribution (different for each mock dataset) at the randomly selected epochs. Then, we have injected two planetary signals with semi-amplitudes $K_{\rm b}$=2.6 and $K_{\rm c}$=3.4 \ms and circular orbits, using the ephemeris derived from the K2 light curve. The internal errors $\sigma_{\rm RV}$ of the additional RVs have been randomly drawn from a normal distribution centred on the mean of the internal errors of the original RV dataset and with $\sigma$ equal to the RMS of the real array $\sigma_{\rm RV}$. In order to avoid the selection of $\sigma_{\rm RV}$ values that are too optimistic and never obtained for the original dataset, we have simulated only internal errors greater than 1 \ms. 
This analysis suggests that, on average, the significance of the retrieved semi-amplitudes of the planetary signals are expected to increase to 4.4$\sigma$ for K2-36\,b and to 5.6$\sigma$ for K2-36\,c, and the same result is obtained for the planet masses. Concerning the densities, their significance increases to 4$\sigma$ ($\rho_{\rm b}$) and to 2.6$\sigma$ ($\rho_{\rm c}$), thus only of a slight amount with respect to the original dataset. Our conclusion is that, under the hypothesis that the structure of the stellar activity signal is preserved several months after the last epoch of the actual observations, a set of 40 additional RVs (which is an affordable amount of measurements to be collected over one season) does not help in improving the precision of the derived planetary parameters enough to significantly improve the theoretical estimates of the planets' bulk composition based on theoretical models.   

\begin{table*}
  \caption[]{Best-fit solutions for the model tested in this work (quasi-periodic GP model) applied to the HARPS-N RV time series extracted with the DRS and TERRA pipelines. Our global model includes two orbital equations (circular and eccentric case). Uncertainties are given as the $16^{\rm th}$ and $84^{\rm th}$ percentiles of the posterior distributions.}
         \label{Table:percentilesgprv}
         \centering
         \small
   \begin{tabular}{cccc}
            \hline
            \noalign{\smallskip}
            Jump parameter     &  Prior & \multicolumn{2}{c}{Best-fit value} \\
                               & & DRS & TERRA \\ 
            \noalign{\smallskip}
            \hline
            \noalign{\smallskip}
            \textbf{Stellar activity GP model} & & & \\ 
            \noalign{\smallskip}
            $h$ [m$\,s^{-1}$] & $\mathcal{U}$(0,30) & 16.3$^{+3.8}_{-2.7}$ & 17.11$^{+3.9}_{-2.8}$ \\  
            \noalign{\smallskip}
            $\lambda$ [days] & $\mathcal{U}$(0,1500) & 131$^{+55}_{-35}$ & 93$^{+22}_{-20}$ \\ 
            \noalign{\smallskip}
            $w$ & $\mathcal{U}$(0,1) & 0.33$^{+0.06}_{-0.05}$ & 0.33$\pm$0.05 \\ 
            \noalign{\smallskip}
            $\theta$ [days] & $\mathcal{U}$(0,20) & 16.99$^{+0.09}_{-0.07}$ & 17.06$\pm$0.08 \\
            \noalign{\smallskip}
            \hline
            \noalign{\smallskip}
            $\sigma_{\rm jit}$ [m$\,s^{-1}$] & $\mathcal{U}$(0,20) & 2.5$\pm$1.0 & 1.6$^{+0.8}_{-0.9}$ \\
            \noalign{\smallskip}
            \hline
            \noalign{\smallskip}
            $\gamma$ [m$\,s^{-1}$] & $\mathcal{U}$(13550,13750) & 13640.7$^{+4.9}_{-5.2}$ & -1.7$^{+5.1}_{-5.2}$ \\
            \noalign{\smallskip}
            \noalign{\smallskip}
            \hline
            \noalign{\smallskip}
            $K_{\rm b}$ [m$\,s^{-1}$] & $\mathcal{U}$(0,10) & 2.1$\pm$0.9 & 2.6$\pm$0.7 \\
            \noalign{\smallskip}
            $K_{\rm c}$ [m$\,s^{-1}$] & $\mathcal{U}$(0,10) & 2.9$\pm$1.1 & 3.4$\pm$1.0 \\

            \noalign{\smallskip}
            \hline
            \noalign{\smallskip}
            \textbf{Derived quantities} \tablefoottext{a} \\
            $m_{\rm b}$ ($\mearth$) & & 3.2$^{+1.4}_{-1.3}$ & 3.9$\pm$1.1 \\
            \noalign{\smallskip} 
            $m_{\rm c}$ ($\mearth$) & & 6.7$^{+2.7}_{-2.6}$ & 7.8$\pm$2.3 \\
            \noalign{\smallskip} 
            $\rho_{\rm b}$ [g$\,cm^{-3}$] & & 5.9$^{+2.9}_{-2.5}$ & 7.2$^{+2.5}_{-2.1}$ \\ 
            \noalign{\smallskip} 
            $\rho_{\rm c}$ [g$\,cm^{-3}$] & & 1.1$^{+0.6}_{-0.5}$ & 1.3$^{+0.7}_{-0.5}$ \\
            \noalign{\smallskip} 
            \hline
            \noalign{\smallskip} 
            \textbf{Bayesian Evidence $\ln\mathcal{Z_{\rm model\: 3}}$} & & -301.45$\pm$0.02 & -295.5$\pm$0.08 \\ 
            \noalign{\smallskip}
             \hline
     \end{tabular}    
     \tablefoot{
     \tablefoottext{a}{Derived quantities from the posterior distributions. We used the following equations (assuming  $M_{\rm s}+m_{\rm p} \cong M_{\rm s}$): $m_{\rm p}\sini \cong$ ($K_{\rm p} \cdot M_{\rm s}^{\frac{2}{3}} \cdot \sqrt{1-e^{2}} \cdot P_{\rm p}^{\frac{1}{3}}) / (2\pi G)^{\frac{1}{3}}$; $a \cong [(M_{\rm s}\cdot G)^{\frac{1}{3}}\cdot P_{\rm p}^{\frac{2}{3}}]/(2\pi)^{\frac{2}{3}} $, where $G$ is the gravitational constant. We set $e$=0 for circular orbits.}
}
\end{table*}

\section{Discussion}
\label{sect:discussion}

In this Section we first use our measurements of the planet parameters to investigate the bulk structure of K2-36\,b and K2-36\,c in the context of the photo-evaporation as the main driver of the evolution of these close-in, low-mass planets. 

\subsection{Planets in the mass-radius diagram}
\begin{figure*}
   \centering
   \includegraphics[width=\hsize]{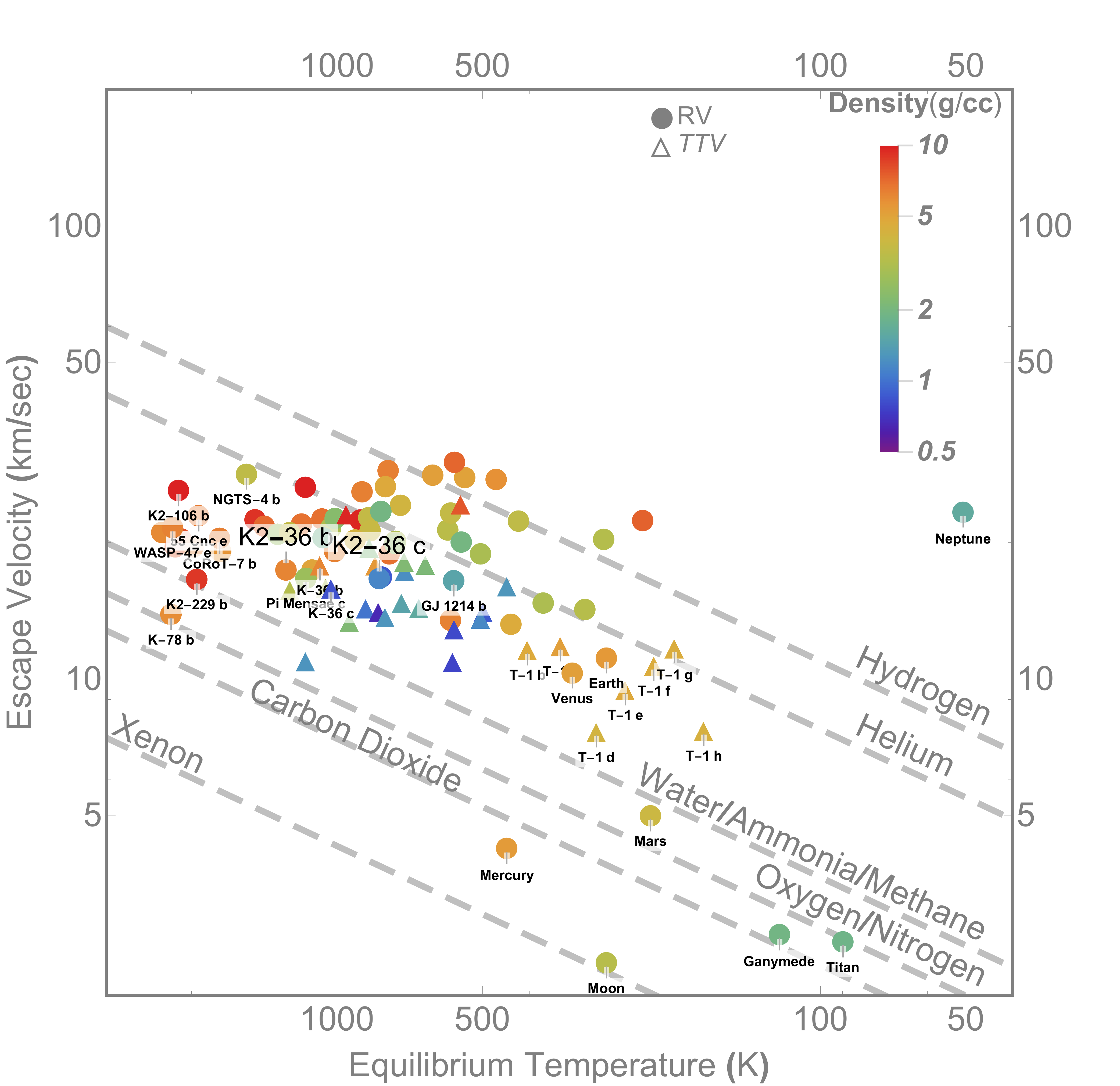}
      \caption{Empirical relations that describe the dependence of the atmospheric composition in exoplanets from their escape velocities and equilibrium temperatures. The plot show the location of the K2-36 planets in the context of some notable small-size exoplanets ($R<4 \rearth$) with measured masses (\textit{K-}: Kepler planets; \textit{T-1}: Trappist-1 planets). A color code is used to distinguish the planets on the basis of their bulk density.}
         \label{fig:atmescape}
\end{figure*}

The exoplanet mass-radius diagram is shown in Fig. \ref{Fig:massradiusdiag}, and it includes the masses of the K2-36 planets derived from the GP analysis of the TERRA dataset.
K2-36\,b is nicely located on the theoretical curve for planets with Earth-like rocky composition, while the Neptune/sub-Neptune size K2-36\,c has a bulk structure compatible with having a H$_{\rm 2}$-dominated gas envelope of $\sim$1$-$2$\%$ planet mass, or a higher mean-molecular-weight envelope with higher mass fraction. Both planets have escape velocity compatible with 20 \kms within 1$\sigma$, as marked in Fig. \ref{Fig:massradiusdiag} by the dashed grey contour corresponding to $m_{\rm p}$/$R_{\rm p}$=3 (in Earth units). Within such a gravitational potential well and considering the young age ($\sim$1 Gyr) of the system, K2-36\,c is consistent with having 1-2$\%$ primordial H2/He-dominated envelope at its corresponding equilibrium temperature (Fig. \ref{fig:atmescape}), or an even less massive H2/He envelope if it has not cooled off from formation yet. Note the mass of K2-36 c is also less than 10 $\mearth$, the critical core mass required for run-away gas accretion \citep{rafikov06}. This may explain why it did not acquire more gas and evolve into a gas giant.


 Both methane CH$_{\rm 4}$ and ammonia NH$_{\rm 3}$ \citep{levi13,levi14,levi17}, and even H$_{\rm 2}$ itself \citep{soubiran15}, can be incorporated into the H$_{\rm 2}$O-reservoir during the initial formation stage, then out-gassed gradually to replenish a primary envelope, or form a secondary envelope, making K2-36\,c a possible example of water-world (Zeng et al. subm.). It is possible that K2-36\,b formed inside the snowline of the system (located at $\sim$1.2 AU, see e.g. \citealt{mulders15}), and K2-36\,c formed farther away beyond the snowline, building up a water-rich core and acquiring almost twice the mass of planet b. In fact, it is expected from cosmic element abundance that just beyond the snowline there is about equal mass available in solids from icy material (including methane clathrates and ammonia hydrates, which both contain H$_{\rm 2}$O) and from rocky material (including primarily Mg-silicates and (Fe,Ni)-metal-alloy), for planet to accrete from, while only rocky material would remain available in solids inside the snowline \citep{lewis72,lewis04}.

\subsection{Comparing K2-36 to Kepler-36}
\begin{figure*}
   \centering
   \includegraphics[width=\hsize]{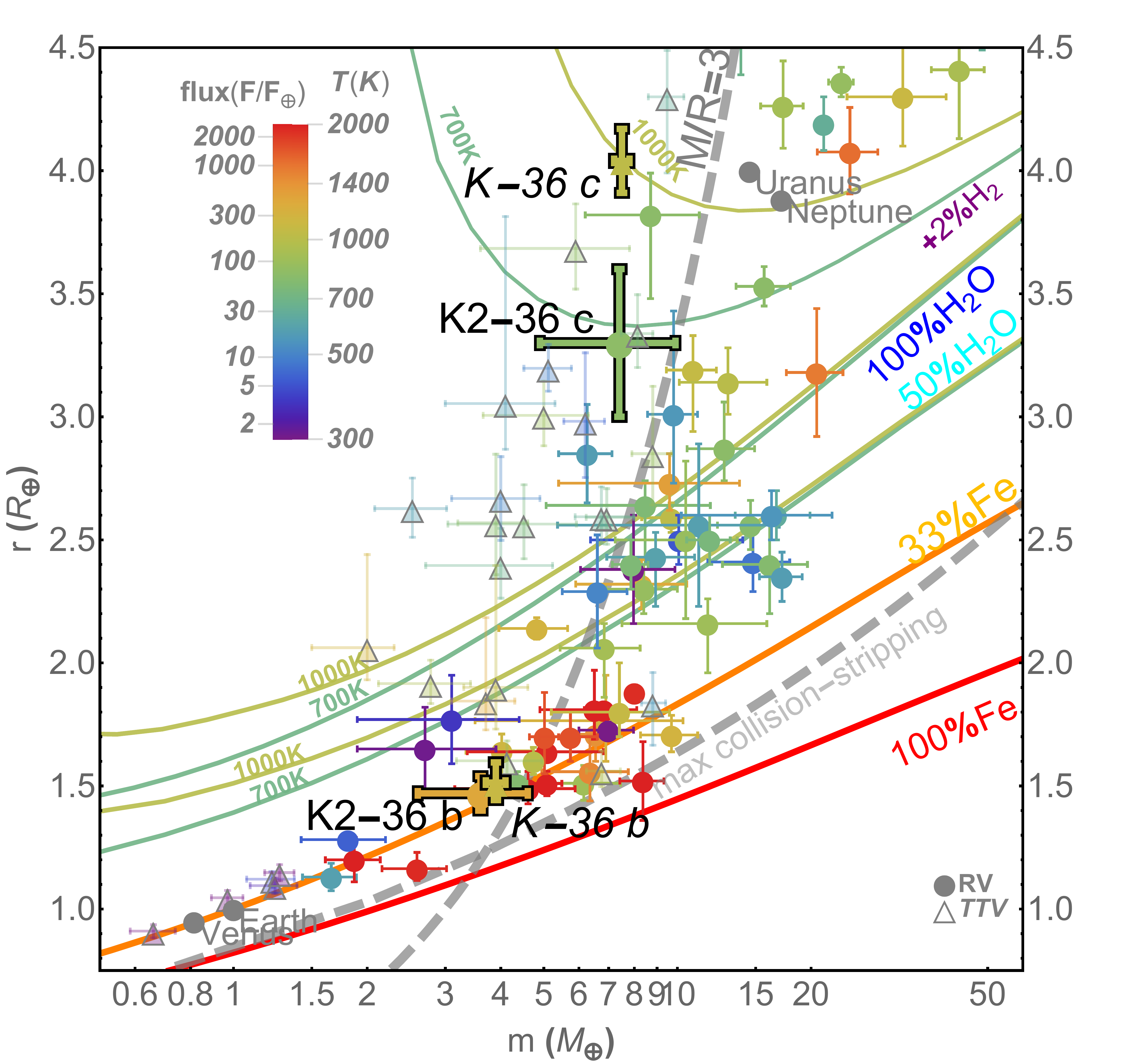}
      \caption{Mass-radius diagram for exoplanets with densities derived with precision better than 35$\%$ (data updated up to 2018 09 14). The planets with masses determined by the radial-velocity method are labelled as circles, and the ones with masses determined by the transit-timing-variation method are labelled as triangles. The color of the data points denotes stellar insolation (see legend in the upper left corner) in the Earth units (expressed as either the amount of stellar bolometric radiation reaching a given area at their orbital distances, assuming negligible orbital eccentricities, normalized to the Earth’s value or surface equilibrium temperatures assuming Earth-like albedo). The plotted masses of the K2-36 planets are those derived from the GP analysis of the TERRA RV dataset. K2-36 planets are represented by filled circles in bold and explicitly labelled. Overplotted are the planets of the Kepler-36 system (filled triangles in bold labelled as K-36), showing an analogy with the K2-36 system as discussed in the text. Two sets of H$_{\rm 2}$O mass-radius theoretical curves (blue - 100 mass$\%$ H$_{\rm 2}$O, cyan - 50 mass$\%$ H$_{\rm 2}$O; cores consist of rock and H$_{\rm 2}$O ice in 1:1 proportion by mass) are calculated for an isothermal fluid/steam envelope at temperatures 700 K and 1000 K, sitting on top of ice VII-layer at the appropriate melting pressure. A set of mass-radius curves (upper portion of the diagram) is calculated for the same temperatures assuming the addition of an isothermal H$_{\rm 2}$-envelope (2$\%$ mass) to the top of the 50 $\%$ mass H$_{\rm 2}$O-rich cores. The dashed curve labelled \textit{maximum collisional stripping} corresponds to the lower bound in radius or upper bound in density that a giant impact can yield for a given planet mass \citep{marcus09}. The dashed curve labelled \textit{M/R=3} corresponds to the contour on the mass-radius diagram where planet mass dividing by planet radius equals 3 in Earth units. It marks the equi-gravitational-potential for the surfaces of planets. Planets residing to the left of this curve are susceptible to atmospheric escape of H$_{\rm 2}$ and Helium over a billion years at their current equilibrium temperatures.}
         \label{Fig:massradiusdiag}
\end{figure*}

An interesting comparison can be made between the K2-36 and Kepler-36 planetary systems. The two-planet system Kepler-36 was discussed by \cite{carter12}, and then studied by \cite{lopezfortney2013}, \cite{quillen13}, \cite{owen16}, \cite{boden18}. Kepler-36 is a 6.8 Gyr moderately-evolved Sun-like star, thus older than K2-36 and with very similar metallicity ([Fe/H]=-0.20$\pm$0.6). Both Kepler-36 planets have very similar masses and radii compared to the corresponding K2-36 planets (see Fig. \ref{Fig:massradiusdiag}. The innermost planets b are consistent with Earth-like rocky composition (see Fig. \ref{Fig:massradiusdiag}). Kepler-36\,c has a larger radius than that of K2-36\,c, by adopting the more recent estimate of \cite{fulton18}\footnote{also \cite{berger18} provided a revised estimate for the radius of Kepler-36\,c, $R=3.689^{+0.165}_{-0.153}$, which is actually very similar to that of \cite{carter12}. By adopting this value instead of that from \cite{fulton18}, Kepler-36\,c and K2-36\,c have their radii compatible within one sigma.}, and lower density.
Kepler-36 planets have different densities and close orbits with periods near the 7:6 mean motion resonance. They both receive nearly half the insolation of K2-36\,b, and have insolation difference much lower than that for K2-36 planets ($[S_{\rm b}-S_{\rm c}]_{\rm Kep-36}\sim 45\, S_{\rm \oplus}$, and $[S_{\rm b}-S_{\rm c}]_{\rm K2-36}\sim 440\, S_{\rm \oplus}$). They are both susceptible to H$_{\rm 2}$-He atmospheric escape considering their escape velocities and current equilibrium temperatures $T_{\rm eq}\geq$900 K. According to their measured densities, Kepler-36\,b has lost any H$_{\rm 2}$-He gaseous envelope, while planet c still appears inflated and consistent with retaining an envelope of H$_{\rm 2}$/He with some percent in mass at the corresponding insolation. The near 7:6 mean motion resonance indicates that they have reached a stable orbital configuration. If photo evaporation has been driving atmospheric loss, its effects could have been steeply accelerated when the host star moved to the MS turn-off point, and it is still eroding the atmosphere of planet c. K2-36 and Kepler-36 planets could have formed in a similar environment, as suggested by the similar host star metallicity, and experienced a similar migration pathway.

\subsection{Comparing K2-36 to WASP-107}
Another interesting comparison is with WASP-107, a one-planet system characterized by a 0.9 $\rjup$ planet with a sub-Saturn mass, and with an orbital period very similar to that of K2-36\,c (5.7 days), transiting a magnetic active K6V star with rotation period very similar to that of K2-36 \citep{anderson17,molcnik17}. Transmission spectroscopy has recently enabled the detection of helium in a likely extended atmosphere of the planet, for which an erosion rate of 0.1-4$\%$ of the total mass per billion years has been calculated \citep{spake18}. WASP-107\,b has an escape velocity ($\geq$20 \kms, or $m_{\rm p}/R_{\rm p}\sim$ 3.6) and a surface equilibrium temperature ($T_{\rm eq}\sim$740 K) not very dissimilar to those of K2-36\,c. If the sub-Neptune K2-36\,c has H$_{\rm 2}$/He as components in its atmosphere, they could currently be escaping at a strong rate, despite intermediate-sized planets (2-4 $\rearth$) presumably having much less H$_{\rm 2}$/He volatiles than WASP-107\,b, which would then be depleted faster. Based on this planet-to-planet comparison, K2-36\,c may be considered suitable for future observations to search for evaporating helium. In fact, by assuming an atmosphere composed of 50$\%$ H$_{2}$ and 50 $\%$He, and the mass and radius estimated above for K2-36c, the atmospheric transmission spectrum signal of the planet can reach values up to about 160 ppm at five scale heights if it has a clear atmosphere. If the planet has a pure He atmosphere that extend 10 scale heights, it might present He absorption with an amplitude of 500 ppm.

\section{Conclusions}
\label{sect:conclusion}
We presented a characterization study of the K2-36 planetary system based on data collected with the HARPS-N spectrograph. According to their size, the planets K2-36\,b and K2-36\,c are located above and below the photo-evaporation valley, and their derived bulk densities ($\rho_{\rm b}$=7.2$^{+2.5}_{-2.1}$ g\,cm$^{-3}$, $\rho_{\rm c}$=1.3$^{+0.7}_{-0.5}$ g\,cm$^{-3}$) indicate that K2-36\,b has an Earth-like rocky composition, while the Neptune/sub-Neptune size K2-36\,c is likely surrounded by a significant gas envelope, not yet evaporated. K2-36 is an ideal laboratory to study the role of the photo-evaporation on the evolution of low-mass, close-in planets, especially for a relatively young ($\sim$1 Gyr) system, as we determined in this work, with the host star having high levels of magnetic activity.

\begin{acknowledgements}
We acknowledge the two anonymous referees for useful comments.
MD acknowledges financial support from Progetto Premiale INAF WOW (\textit{Way to Other Worlds}) and Progetto Premiale 2015 FRONTIERA funding scheme of the Italian Ministry of Education, University, and Research. LM acknowledges financial support from Progetto Premiale 2015 FRONTIERA. ACC acknowledges support from the Science \&\ Technology Facilities Council (STFC) consolidated grant number ST/R000824/1. CAW acknowledges support from the STFC grant ST/P000312/1. 
We thank Dr. B.J.~Fulton and Dr. E.~Petigura for providing us with a preview of the data used in their work. We thank Dr. S.~Engle and J. Kirk for useful discussions.
The HARPS-N project has been funded by the Prodex Program of the Swiss Space Office (SSO), the Harvard University Origins of Life Initiative (HUOLI), the Scottish Universities Physics Alliance (SUPA), the University of Geneva, the Smithsonian Astrophysical Observatory (SAO), and the Italian National Astrophysical Institute (INAF), the University of St Andrews, Queen's University Belfast, and the University of Edinburgh. The research leading to these results received funding from the European Union Seventh Framework Programme (FP7/2007- 2013) under grant agreement number 313014 (ETAEARTH).
Some of this work has been carried out within the framework of the NCCR PlanetS, supported by the Swiss National Science Foundation. 
This study is based upon work supported by the National Aeronautics and Space Administration under grants No. NNX15AC90G and NNX17AB59G issued through the Exoplanets Research Program. This research has made use of the SIMBAD database, operated at CDS, Strasbourg, France, NASA’s Astrophysics Data System and the NASA Exoplanet Archive, which is operated by the California Institute of Technology, under contract with the National Aeronautics and Space Administration under the Exoplanet Exploration Program. Based on observations made with the Italian Telescopio Nazionale Galileo (TNG) operated on the island of La Palma by the Fundacion Galileo Galilei of the INAF (Istituto Nazionale di Astrofisica) at the Spanish Observatorio del Roque de los Muchachos of the Instituto de Astrofisica de Canarias. This paper includes data collected by the K2 mission. Funding for the K2 mission is provided by the NASA Science Mission directorate. Some of the data presented in this paper were obtained from the Mikulski Archive for Space Telescopes (MAST). STScI is operated by the Association of Universities for Research in Astronomy, Inc., under NASA contract NAS5-26555. Support for MAST for non-HST data is provided by the NASA Office of Space Science via grant NNX13AC07G and by other grants and contracts.
This research has also made use of data products from the Wide-field Infrared Survey Explorer, which is a joint project of the University of California, Los Angeles, and the Jet Propulsion Laboratory/California Institute of Technology, funded by the National Aeronautics and Space Administration; of data from the European Space Agency (ESA) mission \textit{Gaia} (https://www.cosmos.esa.int/gaia), processed by the \textit{Gaia} Data Processing and Analysis Consortium (DPAC, https://www.cosmos.esa.int/web/gaia/dpac/consortium). Funding for the DPAC has been provided by national institutions, in particular the institutions participating in the Gaia Multilateral Agreement.

\end{acknowledgements}


\bibliographystyle{aa} 
\bibliography{epic2017} 

\end{document}